\documentclass[apj]{emulateapj}

\usepackage{apjfonts}
\usepackage{amssymb, amsmath}
\usepackage{natbib}
\usepackage{ifthen}
\usepackage{appendix}
\newcommand\degree{\degr}
\newcommand\degrees\degree
\newcommand\vs{\em vs.}

\DeclareSymbolFont{UPM}{U}{eur}{m}{n}
\DeclareMathSymbol{\umu}{0}{UPM}{"16}
\let\oldumu=\umu
\renewcommand\umu{\ifmmode\oldumu\else\math{\oldumu}\fi}
\newcommand\micro{\umu}
\renewcommand\micron{\micro m}
\newcommand\microns \micron

\renewcommand\arcsec[0]{$^{\prime\prime}$}

\let\oldsim=\sim
\renewcommand\sim{\ifmmode\oldsim\else\math{\oldsim}\fi}
\let\oldpm=\pm
\renewcommand\pm{\ifmmode\oldpm\else\math{\oldpm}\fi}
\newcommand\by{\ifmmode\times\else\math{\times}\fi}
\newcommand\ttt[1]{10\sp{#1}}
\newcommand\tttt[1]{\by\ttt{#1}}

\newbox{\wdbox}
\renewcommand\c{\setbox\wdbox=\hbox{,}\hspace{\wd\wdbox}}
\renewcommand\i{\setbox\wdbox=\hbox{i}\hspace{\wd\wdbox}}

\newcount\timect
\newcount\hourct
\newcount\minct
\newcommand\now{\timect=\time \divide\timect by 60
         \hourct=\timect \multiply\hourct by 60
         \minct=\time \advance\minct by -\hourct
         \number\timect:\ifnum \minct < 10 0\fi\number\minct}

\newcommand\mctc{\multicolumn{2}{c}}

\catcode`@=11

\newcommand\comment[1]{}
\newcommand\commenton{\catcode`\%=14}
\newcommand\commentoff{\catcode`\%=12}

\renewcommand\math[1]{$#1$}
\newcommand\mathshifton{\catcode`\$=3}
\newcommand\mathshiftoff{\catcode`\$=12}

\comment{the backslash is necessary}

\comment{alignment tab}

\let\atab=&
\newcommand\atabon{\catcode`\&=4}
\newcommand\ataboff{\catcode`\&=12}

\let\oldmsp=\sp
\let\oldmsb=\sb
\def\sp#1{\ifmmode
           \oldmsp{#1}%
         \else\strut\raise.85ex\hbox{\scriptsize #1}\fi}
\def\sb#1{\ifmmode
           \oldmsb{#1}%
         \else\strut\raise-.54ex\hbox{\scriptsize #1}\fi}
\newbox\@sp
\newbox\@sb
\def\sbp#1#2{\ifmmode%
           \oldmsb{#1}\oldmsp{#2}%
         \else
           \setbox\@sb=\hbox{\sb{#1}}%
           \setbox\@sp=\hbox{\sp{#2}}%
           \rlap{\copy\@sb}\copy\@sp
           \ifdim \wd\@sb >\wd\@sp
             \hskip -\wd\@sp \hskip \wd\@sb
           \fi
        \fi}
\def\msp#1{\ifmmode
           \oldmsp{#1}
         \else \math{\oldmsp{#1}}\fi}
\def\msb#1{\ifmmode
           \oldmsb{#1}
         \else \math{\oldmsb{#1}}\fi}
\def\supon{\catcode`\^=7}
\def\supoff{\catcode`\^=12}
\def\subon{\catcode`\_=8}
\def\suboff{\catcode`\_=12}
\def\supsubon{\supon \subon}
\def\supsuboff{\supoff \suboff}

\newcommand\actcharon{\catcode`\~=13}
\newcommand\actcharoff{\catcode`\~=12}

\newcommand\paramon{\catcode`\#=6}
\newcommand\paramoff{\catcode`\#=12}

\comment{And now to turn us totally on and off...}

\newcommand\reservedcharson{\commenton \mathshifton \atabon \supsubon \actcharon
	\paramon}

\newcommand\reservedcharsoff{\commentoff \mathshiftoff \ataboff
	\supsuboff \actcharoff \paramoff}

\catcode`@=12
\reservedcharsoff

\reservedcharson

\comment{ Must have ONLY ONE of these... trust these macros, they work

}

\newcommand{\squishlist}{
 \begin{list}{$\bullet$}
  { \setlength{\itemsep}{1pt}
     \setlength{\parsep}{0pt}
     \setlength{\topsep}{3pt}
     \setlength{\partopsep}{0pt}
     \setlength{\leftmargin}{2.0em}
     \setlength{\labelwidth}{1.5em}
     \setlength{\labelsep}{0.5em} } }

\newcommand{\squishend}{
  \end{list}  }

\reservedcharson

\actcharon

\shorttitle{Transmission Spectroscopy of the Hot-Jupiter WASP-12b from 0.7 to 5 {\microns}}
\shortauthors{Stevenson {\em et al.}}

\submitted{}

\begin{document}

\title{Transmission Spectroscopy of the Hot-Jupiter WASP-12\lowercase{b} from 0.7 to 5 {\microns}}

\author{Kevin B.\ Stevenson\altaffilmark{1}}
\author{Jacob L.\ Bean\altaffilmark{1}}
\author{Andreas Seifahrt\altaffilmark{1}}
\author{Jean-Michel D\'esert\altaffilmark{2}}
\author{Nikku Madhusudhan\altaffilmark{3,4}}
\author{Marcel Bergmann\altaffilmark{5}}
\author{Laura Kreidberg\altaffilmark{1}}
\author{Derek Homeier\altaffilmark{6}}

\affil{\sp{1}Department of Astronomy and Astrophysics, University of Chicago, 5640 S Ellis Ave, Chicago, IL 60637, USA}
\affil{\sp{2}CASA, Department of Astrophysical and Planetary Sciences, University of Colorado, 389-UCB, Boulder, CO 80309, USA}
\affil{\sp{3}Department of Physics \& Department of Astronomy, Yale University, P.O. Box 208120, New Haven, CT 06520, USA}
\affil{\sp{4}Institute of Astronomy, University of Cambridge, Madingley Road, Cambridge, CB3 0HA, UK}
\affil{\sp{5}National Optical Astronomy Observatory (NOAO), Tucson, AZ 85719, USA}
\affil{\sp{6}Centre de Recherche Astrophysique de Lyon, UMR 5574, CNRS, Universit\'e de Lyon, \'Ecole Normale Sup\'erieure de Lyon, 46 All\'ee d'Italie, F-69364 Lyon Cedex 07, France}

\email{E-mail: kbs@uchicago.edu}

\begin{abstract}

Since the first report of a potentially non-solar carbon-to-oxygen ratio (C/O) in its dayside atmosphere, the highly irradiated exoplanet WASP-12b has been under intense scrutiny and the subject of many follow-up observations.  Additionally, the recent discovery of stellar binary companions \sim1{\arcsec} from WASP-12 has obfuscated interpretation of the observational data.  Here we present new ground-based multi-object transmission-spectroscopy observations of WASP-12b that we acquired over two consecutive nights in the red optical with Gemini-N/GMOS.  After correcting for the influence of WASP-12's stellar companions, we find that these data rule out a cloud-free, H\sb{2} atmosphere with no additional opacity sources.  We detect features in the transmission spectrum that may be attributed to metal oxides (such as TiO and VO) for an O-rich atmosphere or to metal hydrides (such as TiH) for a C-rich atmosphere.  We also reanalyzed NIR transit-spectroscopy observations of WASP-12b from {\em HST/WFC3} and broadband transit photometry from {\em Warm Spitzer}.  We attribute the broad spectral features in the WFC3 data to either H\sb{2}O or CH\sb{4} and HCN for an O-rich or C-rich atmosphere, respectively.  The {\em Spitzer} data suggest shallower transit depths than the models predict at infrared wavelengths, albeit at low statistical significance.  A multi-instrument, broad-wavelength analysis of WASP-12b suggests that the transmission spectrum is well approximated by a simple Rayleigh scattering model with a planet terminator temperature of 1870 {\pm} 130 K.  We conclude that additional high-precision data and isolated spectroscopic measurements of the companion stars are required to place definitive constraints on the composition of WASP-12b's atmosphere.
\end{abstract}
\keywords{planetary systems
--- stars: individual: WASP-12
--- techniques: spectroscopic
}

%%%%%%%%%%%%%%%%%%%%%%%%%%%%%%%%%%%%%%%%%%%%%%%%%%%%%%%%%%%%%%%%%%%%%%%%%%%%%%%
\section{INTRODUCTION}
\label{intro}
%%%%%%%%%%%%%%%%%%%%%%%%%%%%%%%%%%%%%%%%%%%%%%%%%%%%%%%%%%%%%%%%%%%%%%%%%%%%%%%

The advent of ground-based multi-object spectroscopy observations of transiting exoplanets \citep{Bean2010} has opened up the field of atmospheric characterization to more targets due to the availability of additional telescopes and to new wavelengths previously inaccessible using only space-based instruments.

The highly irradiated Jupiter-sized exoplanet WASP-12b \citep{Hebb2009} is currently a target of multiple studies, each working to constrain its atmospheric properties.  In \citet{Madhu2011-wasp12b}, they report the first detection of a planetary atmosphere with a carbon-to-oxygen ratio (C/O) $\ge$ 1.  Using the {\em Spitzer Space Telescope} to observe WASP-12b during secondary eclipse, they find that its dayside atmosphere is enhanced in methane and depleted in water vapor, each by more than two orders of magnitude relative to chemical-equilibrium models with solar abundance. The observed concentrations are consistent with theoretical expectations for an atmosphere with a C/O in excess of unity.  

{\em Spitzer} full-orbit phase observations of WASP-12b at 3.6 and 4.5 {\microns} present conflicting results.  In their best-fit solution, \citet{Cowan2012} report finding transit depths that are inconsistent with model predictions, irrespective of the C/O, eclipse depths that are consistent with previous results \citep{Madhu2011-wasp12b,Campo2011}, and ellipsoidal variations in the 4.5-{\micron} channel only.  However, by fixing the ellipsoidal variations to zero, \citet{Cowan2012} find that the measured transit depths are more consistent with model predictions and that the measured eclipse depths favor a solar C/O and a modest thermal inversion.

The announcement by \citet{Bergfors2013} of a companion star only 1{\arcsec} from WASP-12 has serious implications on the previous results discussed above.  The companion contaminates the measured transit and eclipse depths by upwards of 15\% in the infrared.  Using a spectral type of M0 -- M1, \citet{Crossfield2012} derive wavelength-dependent dilution factors and present corrected transit and eclipse depths for previous analyses.  When combined with their narrow-band, 2.315-{\micron} secondary-eclipse measurement, they find that WASP-12b's emission spectrum is well-approximated by a blackbody and conclude that its photosphere is nearly isothermal.  If true, transmission spectroscopy may be the only method of constraining the atmospheric C/O.  \citet{Bechter2013} have since demonstrated that the companion star is a binary (labeled WASP-12BC) that is physically associated with the primary star WASP-12(A), thus forming a hierarchical triple system.  Using NIR color information, they estimate both companions to be of spectral type M3.

\citet{Swain2013} report on transit and eclipse observations of WASP-12b using {\em Hubble Space Telescope's} (HST) WFC3 instrument.  Using the Bayesian Information Criteria (BIC) as their metric, \citet{Swain2013} find that the best fit to the companion-star-corrected terminator and dayside spectra comes from a pure H\sb{2} atmospheric model with no additional opacity sources.  However, their data cannot rule out more complex (and realistic) models, including one with opacity from the metal hydrides TiH and CrH.  Additionally, they find no evidence of a C/O $\ge$ 1 or a thermal inversion.  \citet{Swain2013} conclude that additional data or detailed modeling is needed to place better constraints on the planet's atmospheric composition.  Two other works also analyzed the WFC3 transmission-spectroscopy data of WASP-12b.  Of note, \citet{Mandell2013} performed an in-depth examination of the band-integrated time series and \citet{Sing2013} explored atmospheric models that include significant opacity from aerosols.

In this paper, we present new ground-based transmission-spectroscopy observations of WASP-12b in the red optical.  The purpose of these observations is to independently constrain the atmospheric C/O by measuring the relative abundance of H\sb{2}O at the terminator.  Additionally, the observations are sensitive to the strong potassium (K) resonance doublet near 770 nm.  The detection of K would limit the presence of hazes in the upper atmosphere.  We also present reanalyses of WASP-12b transmission-spectroscopy observations using {\em HST/WFC3} and broadband-photometry transit observations using {\em Spitzer}, each previously discussed by \citet{Swain2013} and \citet{Cowan2012}, respectively.  

We note that the atmospheric composition at the terminator region does not necessarily need to be consistent with the composition on the dayside.  This is because the transmission spectrum probes a different part of the atmosphere than the emission spectrum.  In addition, \citet{Showman2013} suggest that highly irradiated exoplanets (such as WASP-12b) experience strong thermal forcing that damps planetary-scale waves, thus inhibiting jet formation and efficient circulation at lower altitudes.  This, in turn, inhibits horizontal quenching of molecular abundances to dayside values and can lead to substantial variations in the molecular abundances \citep{Agundez2012}.

Sections \ref{sec:gmosobs}, \ref{sec:wfc3obs}, and \ref{sec:iracobs} discuss observations, data reduction, and light-curve systematics and fits for the Gemini-N/GMOS, {\em HST/WFC3}, and {\em Spitzer/IRAC} data sets, respectively.  In these sections, we also present a new method for modeling spectroscopic light curves and compare results with existing techniques.  In Section~\ref{sec:dilution}, we describe how we account for contamination from the stellar companions WASP-12BC and present corrected transit-depth values with uncertainties.  We present theoretical atmospheric models in Section~\ref{sec:results} and discuss their implication on the planet's C/O.  Finally, we give our conclusions in Section~\ref{sec:concl}.

%%%%%%%%%%%%%%%%%%%%%%%%%%%%%%%%%%%%%%%%%%%%%%%%%%%%%%%%%%%%%%%%%%%%%%%%%%%%%%%
\section{Gemini-N/GMOS OBSERVATIONS AND DATA ANALYSIS}
\label{sec:gmosobs}
%%%%%%%%%%%%%%%%%%%%%%%%%%%%%%%%%%%%%%%%%%%%%%%%%%%%%%%%%%%%%%%%%%%%%%%%%%%%%%%

\begin{deluxetable*}{cccc}
\tabletypesize{\scriptsize}
\tablecolumns{4}
\tablewidth{0pc}
\tablecaption{Observing Log}
\tablehead{
 \colhead{UT Date} &
 \colhead{Exposure Times (s)} &
 \colhead{\# of Exposures} &
 \colhead{Airmass}
}
\startdata
2012 Jan 25 05:32 $\rightarrow$ 12:00 & 140, 180 & 132 & 1.36 $\rightarrow$ 1.02 $\rightarrow$ 1.45 \\
2012 Jan 26 04:56 $\rightarrow$ 12:57 & 180 & 144 & 1.53 $\rightarrow$ 1.02 $\rightarrow$ 1.97 
\enddata
\label{tab:log}
\end{deluxetable*}		

\subsection{Observations}

We observed two transits of WASP-12b using the Gemini-North telescope located atop Mauna Kea, Hawai'i (Program ID GN-2011B-C-1). The Gemini Multi-Object Spectrograph \citep[GMOS,][]{Hook2004} monitored the transits on two consecutive nights in January 2012. A log of the observations is given in Table~\ref{tab:log}.  The conditions were photometric on both nights. The first night was fairly dry, with an estimated precipitable water vapor of 0.7\,mm, while the second night was considerably wetter, with an estimated precipitable water vapor of 1.2\,mm. 

We used the multi-object spectroscopy mode of GMOS with a slit mask to gather time-series spectra of WASP-12 and two other comparison stars. The observations were similar to the multi-object spectroscopy observations pioneered by \citet{Bean2010, Bean2011, Bean2013} and used by \citet{Gibson2013}. The slits in the mask were 12\arcsec\ wide and 30\arcsec\ long. We used the R600\_G5304 grating in first order with a requested central wavelength of 860\,nm. The resulting dispersion was 0.094\,nm\,pixel$^{-1}$ (note the two pixel binning in the dispersion direction, see next paragraph). The OG515\_G0306 filter was used to block light below 515\,nm from higher orders. We obtained spectra of WASP-12 and the brightest comparison star from 719 to 1010\,nm, with two gaps corresponding to the spacing between the three CCDs in the detector mosaic.  The other comparison star was offset from the center of the field of view in the dispersion direction and we obtained spectra from 734 to 1025\,nm for this object.

We recorded the data using the recently-installed GMOS detector array that is populated with e2v deep depletion CCDs. These detectors have substantially improved quantum efficiency and reduced fringing for red optical wavelengths, making them extremely beneficial for our program. At the time of the observations, only a single read mode of the detector array had been characterized. We utilized this mode for our observations, which was the ``slow'' read speed and ``low'' gain (approximately 2.3\,$e^{-}$/ADU) mode using six amplifiers and 2 x 2 binning. We also windowed the detector to read out only the regions around the spectra for the target and two comparison stars. The overhead per exposure was 19\,s, and the duty cycle was better than 88\%. 

For both nights, we observed complete transits without interruption and obtained at least 45\,min of data before and after each transit. All observations on the first night were obtained at airmass less than 1.45, and the observations on the second night were obtained at airmass less than 2.0, with most taken less than 1.5. The GMOS On-Instrument Wavefront Sensor (OIWFS) provided corrections for the deformation of the primary mirror and fast tip-tilt guiding with the secondary mirror.

\subsection{Reduction, Extraction, and Calibration of Spectra}
\label{sec:recs}

Our spectrum reduction, extraction and calibration pipeline is custom software that produces multi-wavelength, systematics-corrected light curves from which we derive wavelength-dependent transit depths with uncertainties.  Previous iterations of the pipeline are described by \citet{Bean2010} and \citet{Bean2011}.  Here we focus our discussion on steps of the data analysis pipeline that differ from these prior versions.

\begin{figure}[tb]
\centering
\includegraphics[width=1.0\linewidth,clip]{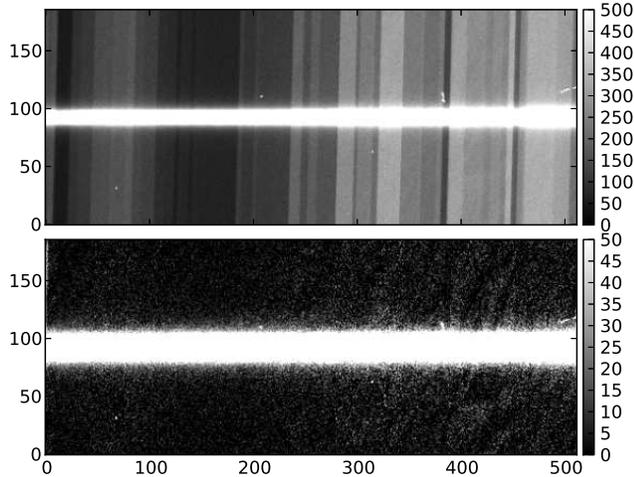}
\caption{\label{fig:w12-background}{
Typical science frame before and after background subtraction.  The axes indicate pixel number in the windowed frame.  The upper panel exhibits a strongly varying background that is misaligned along the spatial direction.  After performing background subtraction as described in the text, the lower panel displays no significant artifacts.  Imprecise background subtraction severely degrades the quality of the light curves.  Note the different grayscales for each panel.  This particular frame is from the \sim950-nm region of the second comparison star in Figure~\ref{fig:wasp12b-spectra} and peaks at \sim6,500 ADU.
}}
\end{figure}

The code reduces the raw science frames by applying the acquired bias and spectroscopic flat frames.  The bias correction is a series of bias frames stacked to form a single master bias frame that is applied uniformly to all of the science frames.  As a test, we also applied the bias overscan regions acquired from and applied to individual science frames and achieved similar results.  We use a CuAr lamp with the OG515 filter and a narrow, 1\arcsec\ slit width from the calibration mask to perform wavelength calibration and to measure and correct for the slit tilt in the reduced data (see Figure~\ref{fig:w12-background}, upper panel).  This effect varies with distance from the center of the optical axis and is due to light distortion in the instrument optics.

%Plot science frame before/after slitshift correction and after background subtraction

\citet{Gibson2013} report seeing fringing at wavelengths >750~nm using Gemini-S/GMOS, but incurred problems using their wide-slit spectroscopic flat fields, so they decided not to apply flat fielding.  We acquired spectroscopic flat frames using both the narrow, 1\arcsec\ slit from the calibration mask and the wide, 12\arcsec\ slit from the science mask to test their effectiveness during the reduction process.  We extracted transmission spectra using both types of frames, fit models to the light curves, and compared the wavelength-varying rms values to those achieved without a flat frame.  Both flat field frames reduce the rms values for wavelengths longer than 900 nm, where we see the effects of fringing, while there is no significant difference in the rms values below 900 nm.  Using the narrow slit, the fringing pattern exhibits sharp peaks with amplitudes in the range of 2--5\%.  Using the wide slit, the fringing pattern undergoes extensive smearing such that the amplitude is <1\%.  In our tests, neither flat field consistently outperformed the other, so we select the wide-slit flat frames for our final analysis because we also acquired the data with that slit.

Standard background-subtraction techniques do not adequately clean up the science frames in the presence of the observed slit tilt.  To achieve the best possible precision in our light curves, we take additional steps when removing the background.  First, we upsample each science frame by a factor of two and correct the detector-spectrum misalignment through interpolation.  Larger upsampling factors did not improve the final results.  Next, we model the sky background by masking out the stellar spectra in the corrected, upsampled science frame and then by fitting a line to each pixel column in the spatial direction.  We generate an out-of-spectra bad-pixel mask by performing a 5$\sigma$ rejection of the residuals along each column.  If a bad pixel is found, it is flagged and we repeat the sky background fitting procedure for that column.  Finally, we interpolate the corrected background frame to its misaligned state, downsample it to its original resolution, then subtract it from the original science frame.  In this way, we can perform precise background subtraction without modifying the science frame through interpolation.  Figure~\ref{fig:w12-background} (lower panel) displays a typical background-subtracted science frame.

\begin{figure}[tb]
\centering
\includegraphics[width=1.0\linewidth,clip]{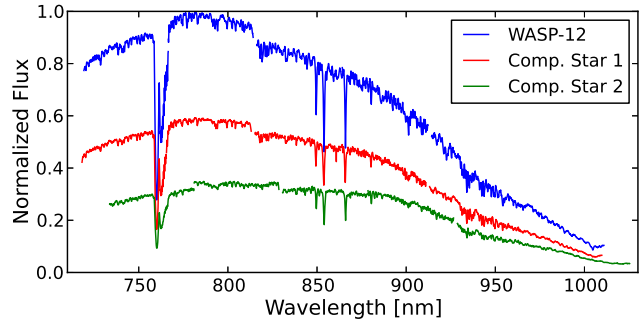}
\caption{\label{fig:wasp12b-spectra}{
Sample spectra of WASP-12 and both comparison stars using the R600\_G5304 grating in the first order.  The two-amplifier read mode with three CCDs outputs six FITS extensions, thus dividing each spectrum into six components.  When correcting for atmospheric fluctuations in our final analysis, we use the brighter of the two comparison stars, which is aligned spectroscopically with WASP-12 on the detectors.
}}
\end{figure}

After performing background subtraction, we execute the optimal extraction routine described by \citet{Horne1986} to produce our final spectra (see Figure~\ref{fig:wasp12b-spectra}).  We cross correlate each spectrum with the first spectrum to measure and correct for its drift over time.  We then generate the wavelength-dependent light curves by binning the spectra into channels with widths of our choosing and summing the results.  During spectral extraction, we do not account for the slight wavelength smearing due to the slit tilt because the maximum tilt is significantly smaller than our bin size (\sim1\%).  For the wavelength-independent (white) light curves, we divide by one of the comparison-star light curves to remove the effects of variable exposure times (if any) and to minimize variations due to fluctuations in Earth's atmosphere.

%Plot sample spectrum

\subsection{Light-Curve Systematics}

Since the multi-object ground-based spectroscopy technique is relatively new, we present an in-depth investigation of the instrument-dependent systematics pertaining to Gemini-N/GMOS observations.  A previous analysis of Gemini-S/GMOS systematics by \citet{Gibson2013} used the Gaussian process method to model unidentified light-curve systematics.  Here we examine correlations between the recorded instrument state values and the light-curve systematics, then derive a physical model that we validate for multiple target/comparison-star combinations.  In addition to the WASP-12 observations, our investigation into light-curve systematics includes two HAT-P-7 data sets.

\begin{figure}[tb]
\centering
\includegraphics[width=1.0\linewidth,clip]{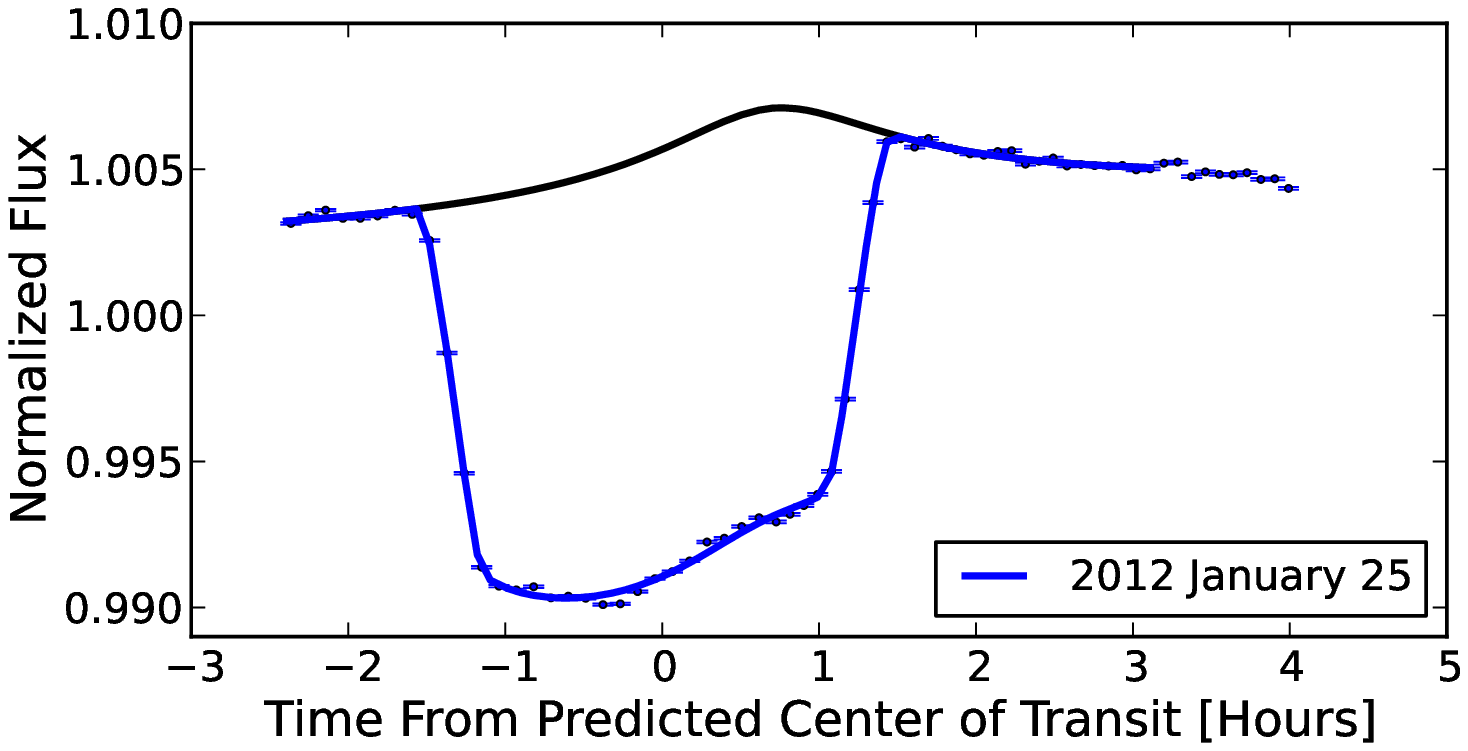}
\includegraphics[width=1.0\columnwidth,clip]{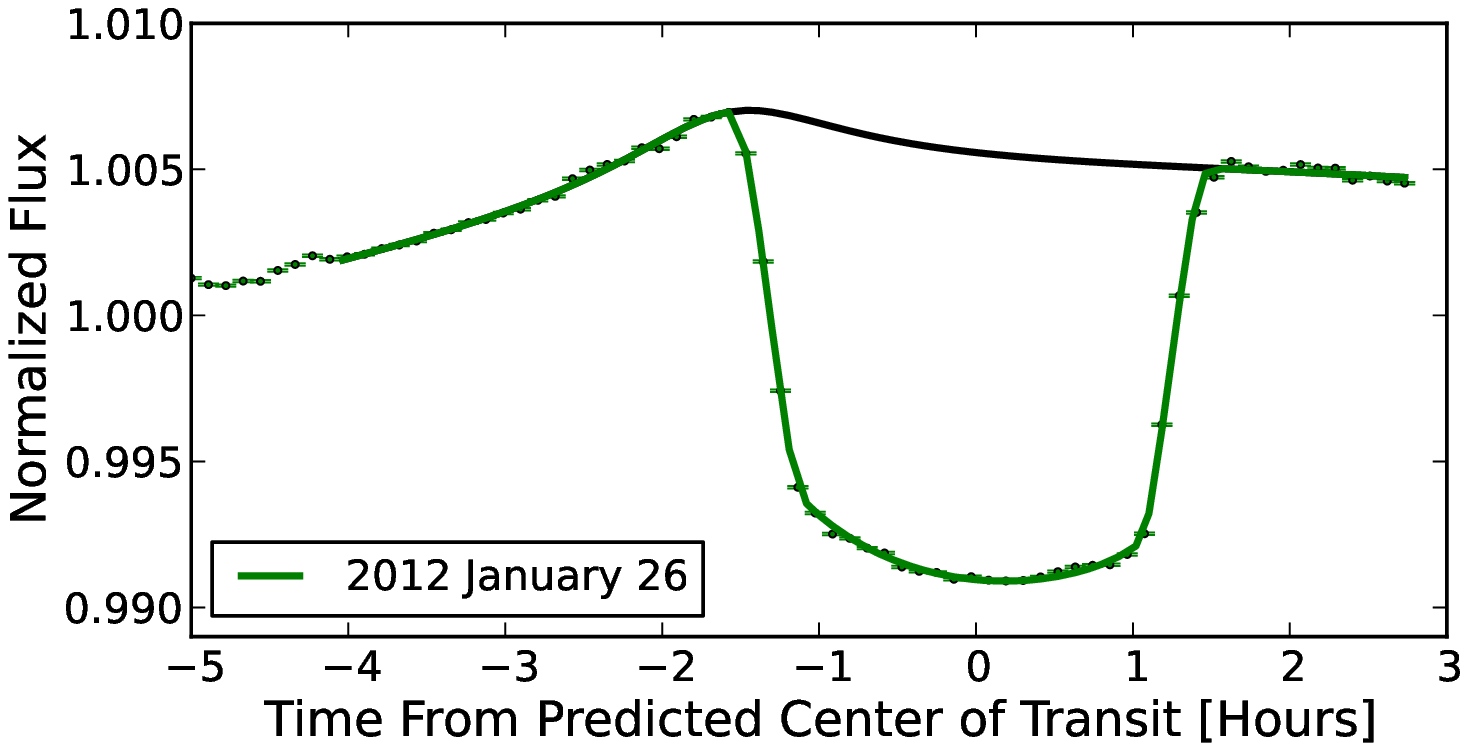}
\includegraphics[width=1.0\linewidth,clip]{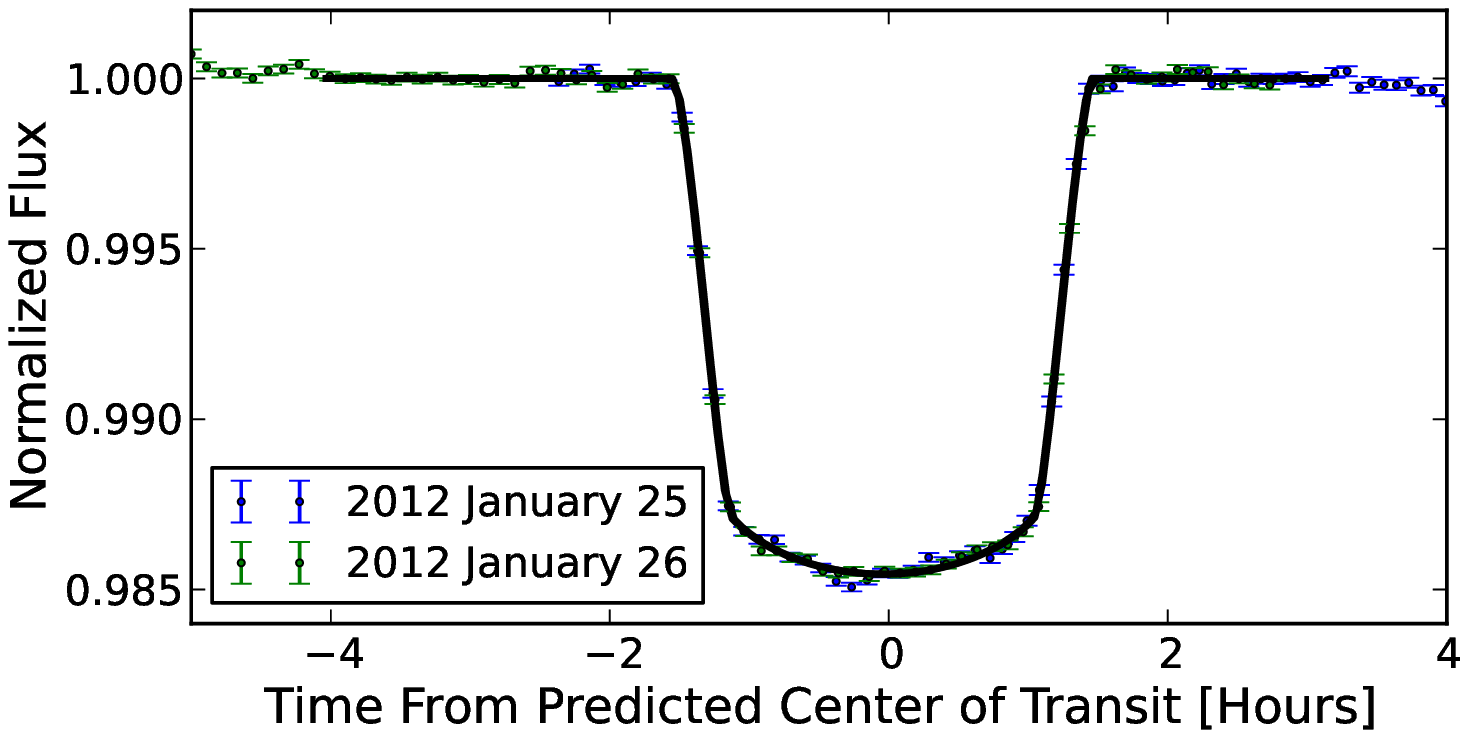}
\caption{\label{fig:wa012b-bin}{
White light curves and best-fit transit models of WASP-12b.  The upper and middle panels depict light curves of WASP-12 from 2012 January 25 and 26, respectively, that are binned in pairs and corrected for atmospheric variations using a comparison star.  We model the instrument-related systematic using Equation \ref{eqn:rotation} and apply the full model given by Equation \ref{eqn:full}.  We do not use flux from 720 -- 765 nm to construct the white light curves because of an anomalous increase in flux towards the end of each observation, as discussed in Section~\ref{sec:fits}.  We exclude points that are far from each transit because the systematic models provide a less-than-ideal fit to these data, which can skew the best-fit transit parameters.  The lower panel presents the normalized, systematics-removed light curves with 1$\sigma$ uncertainties and a best-fit transit model in black.  The residual rms value for each white light curve is 180 ppm and uncertainties are 3.1$\times$ the photon-limit.
}}
\end{figure}

The WASP-12 white light curves in Figure~\ref{fig:wa012b-bin} clearly exhibit a systematic increase in flux towards the midpoint of each observation.  This coincidentally coincides with the zero hour angle for these data, but this chance alignment does not repeat for the HAT-P-7 observations.  After examining the entries in the science header files, we find that the cosine of the Cassegrain Rotator Position Angle (CRPA, denoted $\theta$) plus an offset angle ($\theta\sb{0}$) correlates well with the observed trend in both data sets.  Similarly, the cosine of the parallactic angle of the science target plus a different, comparison-star-dependent offset angle also identically correlates with the observed trend.  For each target star that we analyzed, the angle offset $\theta\sb{0}$ is consistent over both nights; however, this may not be the case if observations are separated by more than a few days.  We also examined the wavelength dependence of the magnitude of this observed systematic.  Longer than $\sim$620~nm, we find that the strength of the systematic is relatively constant and consistent over both observing nights.  Thus, the WASP-12 observations do not require a wavelength-dependent model.  The HAT-P-7 observations, however, extend down to $\sim$510~nm and exhibit a more pronounced instrumental systematic at these bluer wavelengths.

%Plot white light curve with models

The chosen WASP-12 instrument-systematic model component is as follows:
\begin{eqnarray}
\label{eqn:rotation}
S(a,\theta,\theta_0) = 1 + a\cos(\theta + \theta\sb{0}),
\end{eqnarray}
\noindent where $a$ is a multiplicative factor that controls the model component's amplitude, $\theta$ is the time-dependent CRPA as retrieved from the header files, and $\theta_0$ is an angle offset parameter.  As discussed above, we fit the same coefficients to both nights and over all wavelengths.  We also examined model components using other parameters, such as the recorded airmass (or elevation) and the spectrum position on the detector, but find not clear evidence of correlations between these parameters and the instrument systematic for all WASP-12 and HAT-P-7 observations.  We continue to investigate the origins of this instrumental systematic, including ways of predicting the offset term.

The WASP-12 data sets also exhibit time-dependent systematics that we model with linear and quadratic functions.  These are likely unmodeled residuals from the comparison-star spectroscopic corrections.  When fitting both data sets simultaneously, we find that a quadratic function in time with wavelength-dependent free parameters achieves the best fit, as defined by the BIC.  Longer than 770~nm, the measured transit depths do not vary significantly with our choice of ramp model component; transit depths shorter than 770~nm are not well constrained (see Section~\ref{sec:fits}).

\begin{figure}[tb]
\centering
\includegraphics[width=1.0\linewidth,clip]{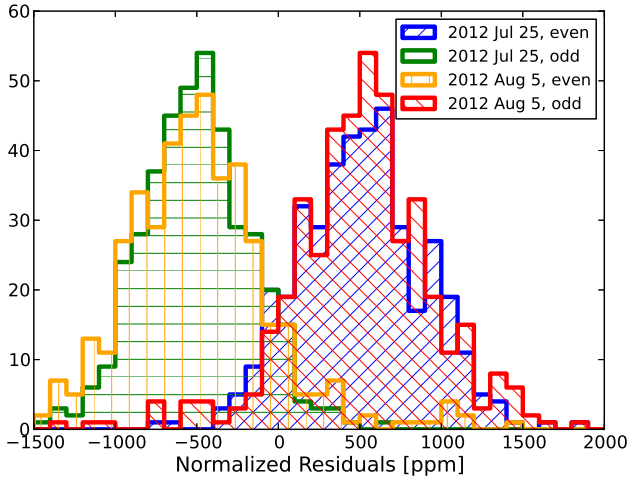}
\caption{\label{fig:hatp7b-hist}{
Histogram of HAT-P-7 light-curve residuals normalized to the out-of-transit flux.  Alternating observations in both data sets are systematically low/high by an average of 500 ppm.  This effect, which we estimate to have an amplitude of 20 -- 30 ppm in the WASP-12 data sets, is not distinguishable in its histograms.  Nonetheless, Method 2 from Section~\ref{sec:nasc} automatically corrects for this systematic when modeling the spectroscopic light curves.
}}
\end{figure}

Our investigation of HAT-P-7 led to the discovery of a second systematic that is not evident in the WASP-12 data because it is significantly weaker in comparison.  As seen in Figure~\ref{fig:hatp7b-hist}, the residuals from alternating frames are systematically offset by $\pm$500 ppm.  This instrumental systematic appears in both HAT-P-7 observations, which predominantly used shorter, 7 -- 9 second exposure times.  We attribute this systematic to the unequal travel times of the GMOS shutter blades, which are known to differ slightly with the direction of motion \citep{Jorgensen2009}.  For comparison, we estimate this effect to have an amplitude of 20 -- 30 ppm in the WASP-12 data sets.  We model this systematic in the HAT-P-7 light curves by applying multiplicative flux offsets to the odd/even frames with the constraint that the product of the offsets is unity.  This minimizes correlations with the system-flux and transit-depth parameters.  Averaging pairs of equal-duration exposures may also be an adequate solution.  In our WASP-12 analysis, the method described in Section~\ref{sec:nasc} automatically corrects for this systematic.

%Plot position offset

%%%%%%%%%%%%%%%%%%%%%%%%%%%%%%%%%%%%%%%%%%%%%%%%%%%%%%%%%%%%%%%%%%%%%%%%%%%%%%%
\subsection{Light-Curve Fits (Method 1)}
\label{sec:fits}
%%%%%%%%%%%%%%%%%%%%%%%%%%%%%%%%%%%%%%%%%%%%%%%%%%%%%%%%%%%%%%%%%%%%%%%%%%%%%%%

We model the light curves using two different techniques then compare the results for consistency.  The first method uses a wavelength-by-wavelength analytic model to describe the systematics:
\begin{eqnarray}
\label{eqn:full}
F(\lambda, t) = F\sb{\rm s}(\lambda)T(\lambda, t)R(\lambda, t)S(a,\theta,\theta_0),
\end{eqnarray}
\noindent where \math{F(\lambda, t)} is the measured flux at wavelength $\lambda$ and time $t$; \math{F\sb{\rm s}(\lambda)} is the wavelength-dependent out-of-transit system flux; \math{T(\lambda, t)} is the primary-transit model component with unity out-of-transit flux; \math{R(\lambda, t) = 1 + r\sb{\lambda,1}(t-t\sb{0}) + r\sb{\lambda,2}(t-t\sb{0})\sp{2}} is the time-dependent ramp model component with a fixed offset, $t\sb{0}$, and wavelength-dependent free parameters, $r\sb{\lambda,1}$ and $r\sb{\lambda,2}$; and $S(a,\theta,\theta_0)$ is the wavelength-independent model component described by Equation \ref{eqn:rotation}.

We fit the light-curve and systematic model components simultaneously using the equations from \citet{Mandel2002} to describe $T(\lambda, t)$ and using free, wavelength-dependent, linear limb-darkening parameters.  The best combination of models is determined by the BIC, which is similar to $\chi\sp{2}$, but accounts for the differing number of free parameters in each model.  We fit the initial models with a least-squares minimizer, rescale all uncertainties to give a reduced $\chi\sp{2}$ of unity, then perform a second minimization.  

In each of the light-curve fits described in Sections \ref{sec:gmosobs} -- \ref{sec:iracobs}, we fix the inclination and semi-major axis to common values (see Table~\ref{tab:transitparams}), which we determine from a joint fit of all of the transit data presented here.  These values are marginally inconsistent with those measured by \citet{Hebb2009} at the 2$\sigma$ level, but are significantly more precise.

\begin{table}[tb]
\centering
\caption{\label{tab:transitparams} 
GMOS White Light Curve Transit Parameters}
\begin{tabular}{ccc}
    \hline
    \hline      
    Parameter               & Value         & Uncertainty   \\
    \hline
    Transit Midpoints (MJD\tablenotemark{a}) 
                            & 5951.83534    & 0.00011       \\
                            & 5952.92720    & 0.00010       \\
    $R\sb{P}/R\sb{*}$       & 0.11713       & 0.00019       \\
    $\cos i$                & 0.164         & 0.006         \\
    $a/R\sb{*}$             & 2.908         & 0.020         \\
    Limb-Darkening Coef.    & 0.265         & 0.009         \\
    \hline
\end{tabular}
\tablenotetext{1}{MJD = BJD\sb{TDB} - 2,450,000}
\end{table}

We estimate errors using both the residual permutation method and a differential-evolution Markov-Chain Monte Carlo (DE-MCMC) algorithm, but use the former for our final GMOS values.  We also investigate the effects of correlated noise using the wavelet analysis described by \citet{Carter2009b}.  The code can run any combination of models and can fit multiple events (nights) simultaneously while sharing parameters between light-curve models.  For the GMOS data, we apply 10,000 random permutations to the time-ordered residuals from each night.  This is possible because the data from one night are independent of data from the other.  At each step, we add the shifted residuals from both nights to the best-fit models from the actual data and calculate a new best fit.

\begin{figure*}[tb]
\centering
\includegraphics[width=0.49\linewidth,clip]{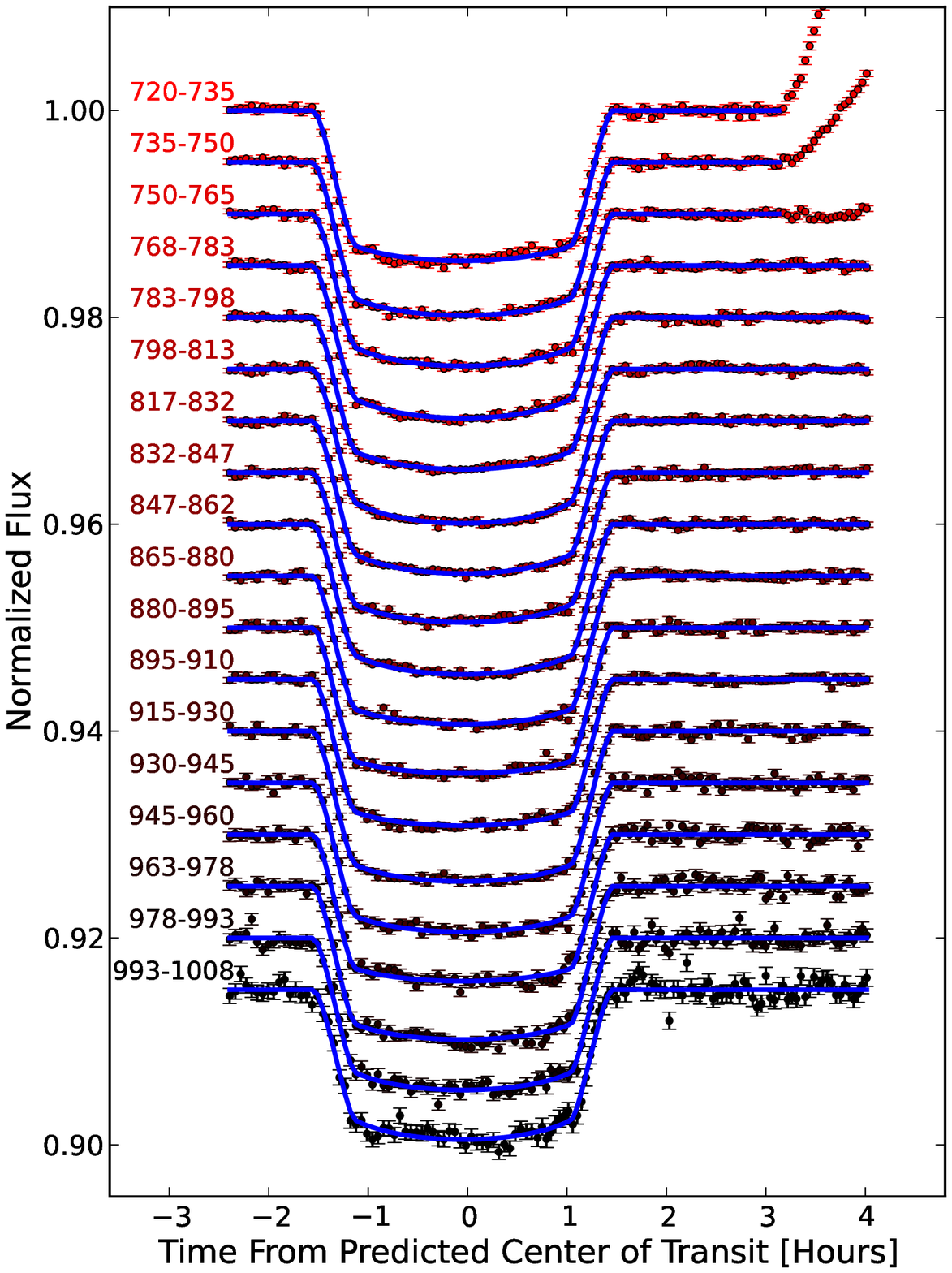}
\includegraphics[width=0.49\linewidth,clip]{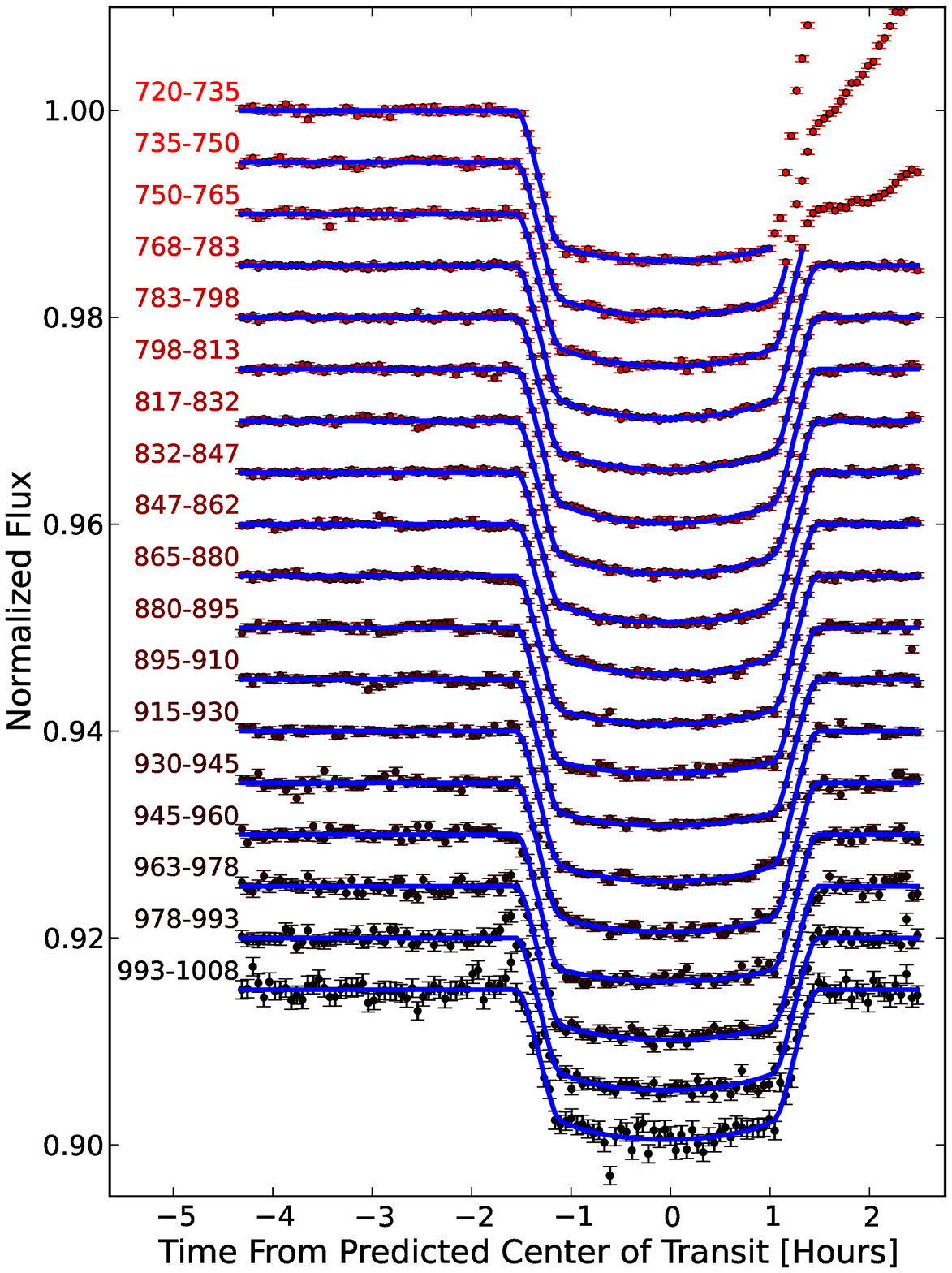}
\caption{\label{fig:wa012b-lc1}{
WASP-12b spectroscopic light curves from 2012 January 25 (left panel) and 2012 January 26 (right panel) using Gemini-N's GMOS instrument.  The methods described in Section~\ref{sec:nasc} produced these results, which are normalized to the system flux and shifted vertically for ease of comparison.  The blue lines are best-fit models and the error bars are 1\math{\sigma} uncertainties.  The wavelength range for each of the 18 channels is specified in nm.  The apparent increase in flux near the end of each night and below 765~nm is the result of a decrease in measured flux from the comparison star.  The residual rms values range from 220 to 760 ppm in the left panel and from 170 to 870 ppm in the right panel.
}}
\end{figure*}

We divide each spectrum into 18 channels, each 15~nm in width, with three channels per detector.  The three bluest channels exhibit an unexplained increase in flux that we do not model (see Figure~\ref{fig:wa012b-lc1}).  The anomaly occurs on both nights at roughly the same local time ($\sim$01:30 HST) and while the objects are still relatively high in the sky (elevation $\sim$ 55\degr).  Since we do not see this effect when applying atmospheric corrections using the fainter comparison star, we conclude that the origin is a wavelength-dependent decrease in measured flux from the brighter comparison star, rather than an increase in measured flux from WASP-12, and is probably due to time-dependent vignetting.

%Plot colored light curves with models

%%%%%%%%%%%%%%%%%%%%%%%%%%%%%%%%%%%%%%%%%%%%%%%%%%%%%%%%%%%%%%%%%%%%%%%%%%%%%%%
\subsection{Light-Curve Fits (Method 2)}
%Non-Analytic Models of Systematics
\label{sec:nasc}
%%%%%%%%%%%%%%%%%%%%%%%%%%%%%%%%%%%%%%%%%%%%%%%%%%%%%%%%%%%%%%%%%%%%%%%%%%%%%%%

Here we introduce a different, independent technique (called {\tt Divide-White}) to model the wavelength-dependent (spectroscopic) light curves, without making any prior assumptions about the form of the systematics, by taking advantage of information within the wavelength-independent (white) light curves.  With this method, the residual rms values are smaller in all channels.  It achieves nearly identical best-fit relative transit depths as the systematics-modeled analysis described in Section~\ref{sec:fits}, but produces noticeably smaller uncertainties longer than 960 nm where there is significantly less flux.

We begin with a brief synopsis of our newly developed method, followed by a detailed explanation:
\begin{enumerate}
\item Model the wavelength-independent (white) light curve as described in Section~\ref{sec:fits}.
\item Use the best-fit parameters to construct a white transit model, $T\sb{white}(t)$.
\item Divide the target star's white light curve by $T\sb{white}(t)$ and normalize the result to derive a non-analytic model of the wavelength-independent systematics, $Z\sb{white}(t)$.
\item Divide the comparison stars' spectroscopic light curves by their respective white light curves, add the comparison-star spectroscopic corrections in each channel if using multiple comparison stars, then normalize the results in each channel to derive non-analytic models of the wavelength-dependent systematics, $Z(\lambda,t)$.
\item Combine the systematics models with wavelength-dependent transit models to construct the final light-curve models, $F(\lambda, t) = F\sb{\rm s}(\lambda)T(\lambda, t)Z\sb{white}(t)Z(\lambda,t)$.  If necessary, append additional terms to model residual systematics.
\item Compare $F(\lambda, t)$ to the data, determine the best-fit solution, and estimate uncertainties.
\end{enumerate}

Much in the same way a comparison star is used to correct measured flux due to atmospheric fluctuations, a white light curve can correct wavelength-independent flux variations in the spectroscopic light curves.  The advantage of this method over using a comparison star is that light from the target and comparison stars is not guaranteed to have traveled through the same column of air.  Additionally, stellar variability in either the target or comparison star can produce a less-than-ideal correction and may even be a significant source of noise.  

Our initial goal is to determine the transit parameters that best describe the white light curve.  This is accomplished by first dividing the target star's white light curve by one or more comparison stars' white light curves.  We then simultaneously model the transit and any significant systematics in the data.  At this point, we are not concerned with estimating uncertainties, so a simple least-squares minimization will suffice.  Also, it is not necessary to model the more subtle systematics that do not impact the measured transit parameters, such as the one discussed in Figure~\ref{fig:hatp7b-hist}.  This is because the transit parameters are the only pieces of information we retain from this step.  We use the best-fit parameters (see Table~\ref{tab:transitparams}) to construct a white transit model, $T\sb{white}(t)$.

Our next goal is to construct non-analytic models of the light-curve systematics.  We readily obtain a model for the wavelength-independent systematics, $Z\sb{white}(t)$, by dividing the transit star's white light curve by the white transit model, $T\sb{white}(t)$.  We normalize the transit-removed results to unity to maintain the physical significance of our system flux parameters, $F\sb{\rm s}(\lambda)$, but this is not a necessary step.  We derive models for the wavelength-dependent systematics, $Z(\lambda,t)$, using information from one or more comparison stars.  For each comparison star, we divide each channel's spectroscopic light curve by its respective white light curve.  For each channel, it is possible to add the result from each comparison star to assemble a more precise estimate of $Z(\lambda,t)$.  Some spectroscopic channels from a comparison star do not behave as expected (see Figure~\ref{fig:wa012b-lc1} and corresponding discussion in Section~\ref{sec:fits}), so we omit those channels from that comparison star when constructing $Z(\lambda,t)$ and rely solely on the spectroscopic correction from the other comparison star.  Again, we normalize $Z(\lambda,t)$ to unity in each channel.

\begin{figure}[tb]
\centering
\includegraphics[width=1.0\linewidth,clip]{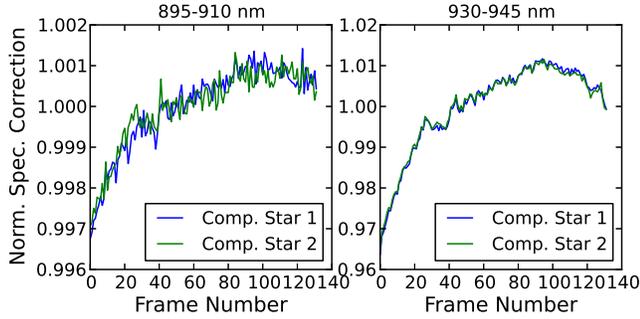}
\caption{\label{fig:refSpecLC}{
Normalized spectroscopic corrections from two comparison stars from 895 to 910 nm (left) and 930 to 945 nm (right).  Although the first comparison star is $\sim$0.8 magnitudes brighter than the second, both produce similar corrections.  The correction magnitudes vary over the channels, from 0.4\% -- 4\%, and are most significant in regions of strong atmospheric absorption due to the presence of H\sb{2}O or O\sb{2}.
}}
\end{figure}

One implication of this method is that, in order to achieve the most precise spectroscopic correction without adversely decreasing the observing efficiency, we are no longer constrained to using comparison stars of similar magnitude to that of the target star (see Figure~\ref{fig:refSpecLC}).  By combining the derived spectroscopic corrections from multiple, fainter comparison stars, we can theoretically achieve the same precision to that of a single, brighter comparison star.  This method opens up the possibility of observing bright (<8 mag), nearby exoplanet host stars that do not have a comparison star of similar magnitude within several arcminutes.

%Plot two (i=11,13) spectroscopic corrections for both comparison stars.

We construct the final light-curve model as follows:
\begin{eqnarray}
\label{eqn:nexis}
F(\lambda, t) = F\sb{\rm s}(\lambda)T(\lambda, t)R(\lambda, t)Z\sb{white}(t)Z(\lambda,t),
\end{eqnarray}
\noindent where $R(\lambda, t)$ is a time-dependent ramp model component with wavelength-dependent free parameters.  In this case, we use $R(\lambda, t)$ to account for residual systematics not fully accounted for by the non-analytic models.  We construct $Z\sb{white}(t)$ using channels with wavelengths >765 nm and construct $Z(\lambda,t)$ using the brighter of the two companion stars.

\begin{figure}[tb]
\centering
\includegraphics[width=1.0\linewidth,clip]{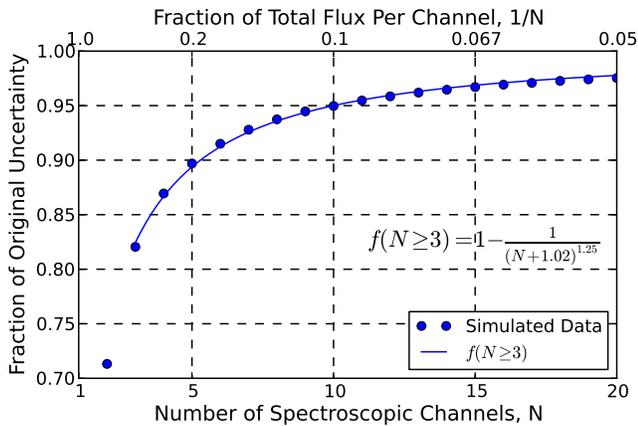}
\caption{\label{fig:correlations}{
Results from simulated data to determine the level of correlation using the technique described in Section~\ref{sec:nasc}.  We plot the ratio of the standard deviations of the spectroscopic channels before and after dividing by the normalized white light curve, $Z\sb{white}(t)$, {\vs} $N$, the number of spectroscopic channels.  As we increase the number of channels, thus decreasing the fraction of contributed flux per channel in constructing $Z\sb{white}(t)$, we expect the ratio of standard deviations to approach unity and the correlation level to approach zero.  We also fit the simulated data with the function $f(N)$, which is valid for $N\geq3$.
}}
\end{figure}

Because we construct $Z\sb{white}(t)$ from the individual spectroscopic channels, neighboring points in the residuals are now slightly correlated.  To determine the level at which we may have reduced the random scatter of points (due to correlations), we simulate $N$ flat spectroscopic light curves with Poisson noise, sum their values to create a white light curve, divide each spectroscopic light curve by the normalized white light curve, then compare the resulting standard deviation of the spectroscopic points to the known standard deviation from the Poisson noise.  Figure~\ref{fig:correlations} depicts the results by plotting the ratio of the standard deviations {\vs}~the number of spectroscopic channels.  Based on these simulations, we conclude that using ten evenly-weighted spectroscopic channels to assemble the white light curve reduces the scatter by \sim5\%; with five channels, the scatter is reduced by \sim10\%.  For comparison, the GMOS and WFC3 data (see Section~\ref{sec:wfc3obs}) use 18 and 11 channels, respectively.

\begin{figure}[tb]
\centering
\includegraphics[width=1.0\linewidth,clip]{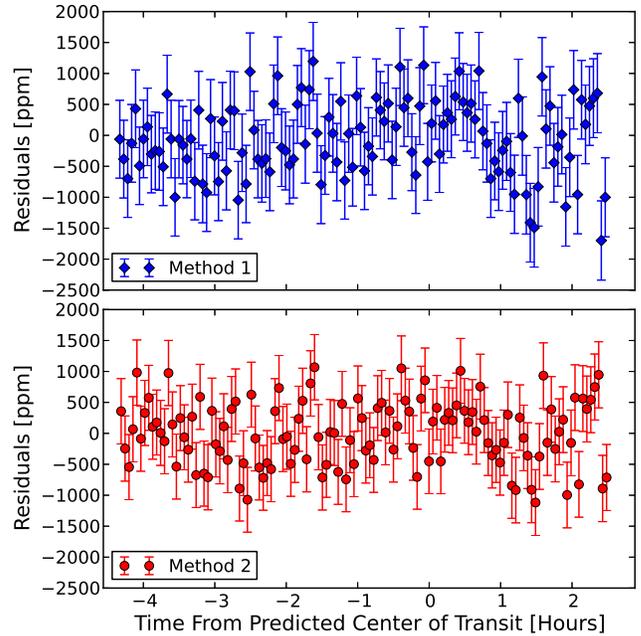}
\caption{\label{fig:CompareResiduals}{
Light-curve residuals from 963 to 978 nm using two different modeling techniques.  Method 1 (top panel) uses analytic models to fit the systematics (see Section~\ref{sec:fits} for details) and has an rms of 630 ppm.  Method 2 (bottom panel) uses information in the white light curve to model the wavelength-independent systematics and has an rms of 520 ppm.  The fact that two different modeling techniques produce nearly identical results adds confidence to our measured transit depths in general and to the observed feature at these wavelengths in particular (discussed in Section~\ref{sec:results}).
}}
\end{figure}

We evaluate the effectiveness of each method by comparing their residual rms values.  For each channel, we find that the {\tt Divide-White} method consistently outperforms the standard method from Section~\ref{sec:fits}.  As an example, Figure~\ref{fig:CompareResiduals} compares the residuals from each method in one channel.  Their average point-to-point difference in this example is less than 0.4$\sigma$ and both sets of residuals clearly depict the same correlated noise.  Upon reviewing all of the channels over both nights, we conclude that both methods produce visually comparable results; however, we find that our second method is more precise.

%Compare 970.5 nm LC residuals from both methods.

We compare the accuracy of both methods by testing them against a fictitious transit signal of known depth.  First, we remove the existing transit signal in the 2012 January 26 dataset by dividing the spectroscopic light curves by their best-fit transit models.  Next, we inject an artificial signal into each channel and then find the best-fit solution with each technique.  By using the same transit depth and limb darkening values at all wavelengths, we are guaranteed to know the white light curve transit parameters with absolute certainty and, as such, can evaluate the effects of improperly fitting the white light curve.  

\begin{figure}[tb]
\centering
\includegraphics[width=1.0\linewidth,clip]{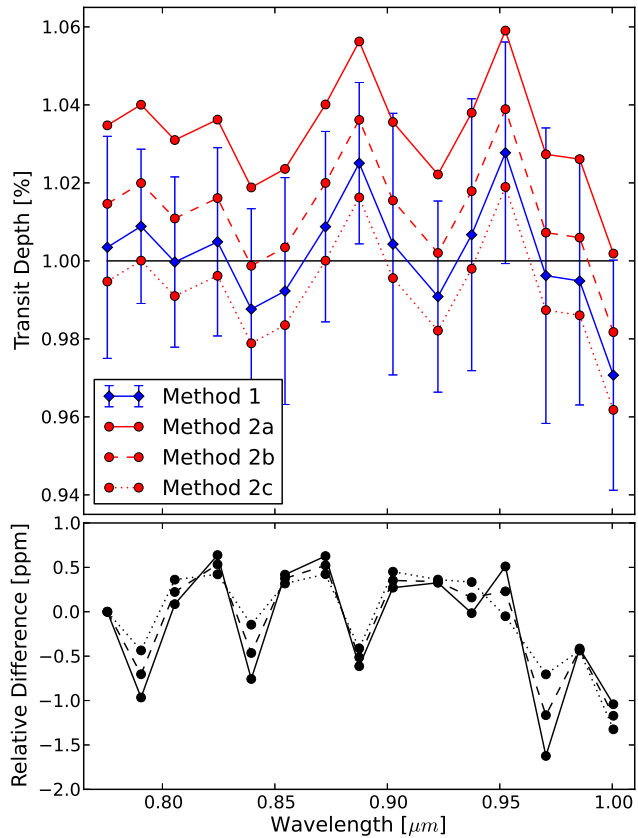}
\caption{\label{fig:fakeTransit}{
Best-fit transit depths and relative differences using a fictitious transit signal.  In the top panel, the solid black line depicts the injected signal ($\delta(\lambda) = 1\%$) at all wavelengths.  The solid blue line (with diamond symbols and 1$\sigma$ uncertainties) indicates the best fit using Method 1.  In red, we show results from Method 2 using three different white light curve transit depths: 1.02 (a), 1.00 (b), and 0.98 \% (c).  The spectroscopic depths are positively correlated with the white depth; therefore, it is important to accurately determine the white depth with this method.  The best-fit transit depths for Methods 1 and 2b are consistent with each other and the injected signal.  The bottom panel plots the relative differences in transit depths between Method 1 and the three versions of Method 2.  Unlike the absolute depths, the relative transit depths are all consistent to within a few ppm.
}}
\end{figure}

In the top panel of Figure~\ref{fig:fakeTransit}, we plot best-fit transit depths for Method 1 and three different white light curve depths using Method 2, all using a fixed linear limb-darkening parameter.  The results show that both methods find the correct transit depths to within 1$\sigma$ and, thus, are also consistent with each other.  Additionally, we note that the spectroscopic light curve depths are positively correlated with the white light curve depth in a one-to-one correspondence.  In the bottom panel of Figure~\ref{fig:fakeTransit}, we show that the relative spectroscopic depths are all consistent to within a couple ppm, regardless of the method used.  Therefore, when using Method 2, we conclude that it is important to correctly determine the white light curve depth to obtain accurate absolute spectroscopic depths; however, the same is not true for the relative spectroscopic depths, which are predominantly independent of the white light curve depth.

In tests where we fit for the limb-darkening, the best-fit transit depths from Method 2 remain unchanged; however, for Method 1, all of the transit depths shift upward by nearly 1$\sigma$.  Fitting for the limb darkening in these tests also increases the relative differences between the methods by a couple ppm.  For the remainder of our analysis, we use Method 2 ({\tt Divide-White}).

%Plot results from injecting a fake signal.

\begin{figure}[tb]
\centering
\includegraphics[width=1.0\linewidth,clip]{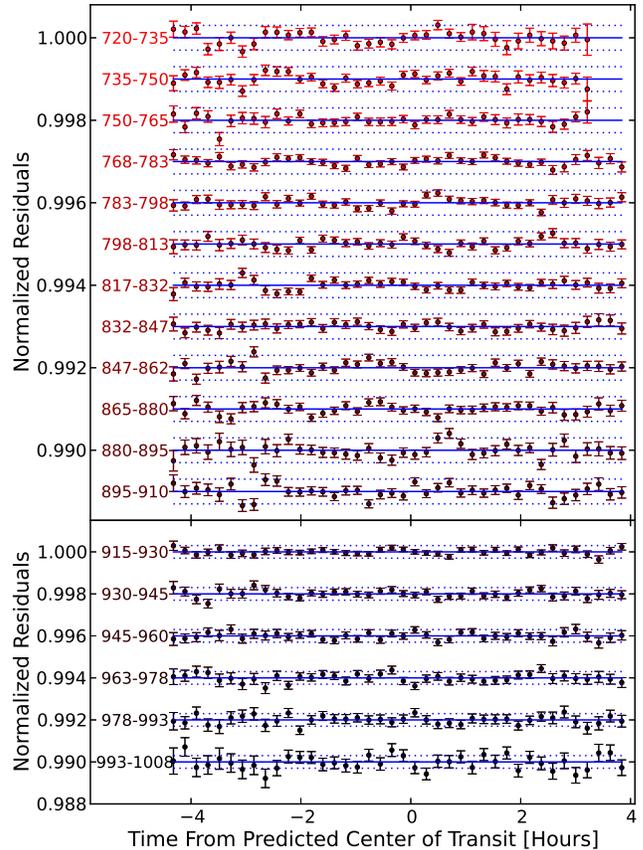}
\caption{\label{fig:Residuals}{
Binned light-curve residuals from the Gemini-N/GMOS observations.
We combine residuals from both transit observations using $\sim$11-minute bins (colored points with 1$\sigma$ uncertainties).  For reference, the solid blue lines indicate the zero levels and the dotted blue lines indicate {\pm}300 ppm.  The wavelength range for each of the 18 channels is specified in nm.  The top panel uses a smaller spacing relative to the bottom panel to more easily distinguish variations in the residual points.
}}
\end{figure}

\begin{figure}[tb]
\centering
\includegraphics[width=1.0\linewidth,clip]{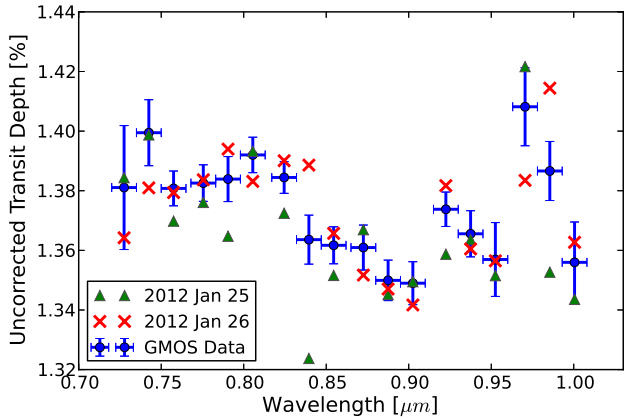}
\caption{\label{fig:gmosspec}{
WASP-12b uncorrected transmission spectrum (with respect to contribution from WASP-12's stellar companion) using Gemini-N's GMOS instrument.  Green triangles and red crosses represent best-fit transit depths from individual nights 2012 January 25 and 2012 January 26, respectively.  Blue circles with 1$\sigma$ uncertainties depict the results of a joint fit with shared parameters.  In 16 of the 18 channels, we obtain consistent results (to within 2$\sigma$) over both nights.
}}
\end{figure}

Using the observational data, Figure~\ref{fig:Residuals} depicts binned residuals of the 18 spectroscopic channels, obtained from a joint fit of both transit observations.  The spectroscopic residuals show no significant deviations from the best-fit models.  Table~\ref{tab:gmosrms} gives the residual rms values and compares those values to what we expect in the limit of Poisson-distributed noise.  The average multiplicative factor is \sim2.1, which is consistent with previous ground-based data.  Due to the averaging of wavelength-dependent effects in the white light curves, some spectroscopic channels achieve comparable rms values to the white light curves; all channels exhibit scatter that is closer to the photon limit.  In Figure~\ref{fig:gmosspec}, we plot the transmission spectrum of WASP-12b prior to correcting for the presence of WASP-12's stellar companion.  With access to two transit events, we confirm that both nights produce consistent results.  In Section~\ref{sec:results}, we present the corrected transmission spectrum and discuss constraints on the atmospheric composition.

%Plot binned residuals from both nights.

\begin{table}[tb]
\centering
\caption{\label{tab:gmosrms} 
GMOS Precision}
\begin{tabular}{ccccc}
    \hline
    \hline      
                & \mctc{2012 January 25}    & \mctc{2012 January 26}    \\
    Wavelength  & rms   & $\times$Photon Limit
                                            & rms   & $\times$Photon Limit      \\
    (nm)        & (ppm)         &           & (ppm)         &       \\
    \hline
    720--735    & 350           & 2.8       & 270           & 2.3   \\
    735--750    & 270           & 2.3       & 250           & 2.2   \\
    750--765    & 240           & 1.9       & 270           & 2.3   \\
    768--783    & 250           & 2.2       & 170           & 1.6   \\
    783--798    & 230           & 2.0       & 190           & 1.8   \\
    798--813    & 230           & 1.9       & 240           & 2.2   \\
    817--832    & 220           & 1.8       & 220           & 2.0   \\
    832--847    & 280           & 2.3       & 230           & 2.0   \\
    847--862    & 230           & 1.8       & 240           & 2.0   \\
    865--880    & 280           & 2.1       & 240           & 1.9   \\
    880--895    & 310           & 2.3       & 350           & 2.7   \\
    895--910    & 290           & 1.9       & 280           & 2.0   \\
    915--930    & 320           & 1.9       & 310           & 1.9   \\
    930--945    & 390           & 1.9       & 420           & 2.2   \\
    945--960    & 460           & 2.1       & 430           & 2.0   \\
    963--978    & 480           & 2.0       & 520           & 2.3   \\
    978--993    & 580           & 2.1       & 580           & 2.2   \\
    993--1008   & 760           & 2.2       & 870           & 2.6   \\
    \hline
\end{tabular}
\end{table}

\clearpage

%%%%%%%%%%%%%%%%%%%%%%%%%%%%%%%%%%%%%%%%%%%%%%%%%%%%%%%%%%%%%%%%%%%%%%%%%%%%%%%
\section{HST/WFC3 OBSERVATIONS AND DATA ANALYSIS}
\label{sec:wfc3obs}
%%%%%%%%%%%%%%%%%%%%%%%%%%%%%%%%%%%%%%%%%%%%%%%%%%%%%%%%%%%%%%%%%%%%%%%%%%%%%%%

\subsection{Observations}

\begin{figure}[tb]
\centering
\includegraphics[width=1.0\linewidth,clip]{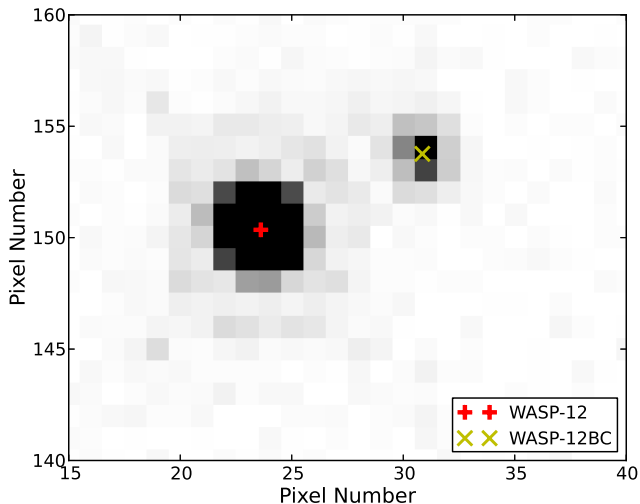}
\caption{\label{fig:wa012b-hst-phot}{
{\em HST} photometric observation of WASP-12 on 2011 April 12.  The binary WASP-12BC (marked with an `x') clearly distinguishes itself from WASP-12 (marked with a `+').  This allows us to directly measure the companion's dilution factor at 1.32 {\microns} (see Section~\ref{sec:dilution}).  We use stellar atmospheric models and the measured offset in the dispersion direction (along the abscissa) to estimate the dilution factor at other wavelengths and correct the measured transit depths for WASP-12b.  Using a plate scale of 0.135$\times$0.121~\arcsec/pixel \citep{wfc3_ihb}, we determine the stars' separation to be 1.061 {\pm} 0.002{\arcsec}, which is in good agreement with \citet{Crossfield2012} and \citet{Bechter2013}.
}}
\end{figure}

{\em HST} observed WASP-12b in staring mode (not spatial scan) during its primary transit on 2011 April 12 \citep[Program number 12230, PI Mark Swain, ][]{Swain2013}.  The WFC3 instrument utilized its G141 GRISM to acquire spectra from 1.1 to 1.7 {\microns} over five HST orbits.  It also acquired a photometric image of WASP-12 and the nearby companion star (see Figure~\ref{fig:wa012b-hst-phot}) using the F132N filter.  We estimate their separation to be 1.061 {\pm} 0.002{\arcsec}, which is in good agreement with \citet{Crossfield2012} and \citet{Bechter2013}.

\subsection{Reduction, Extraction, and Calibration of Spectra}

The steps for reducing and extracting the WFC3 spectra are similar to those described in Section~\ref{sec:recs}, with any differences discussed below.  We use the calibrated ``\_flt'' frames provided by the STScI Archive.  To calculate the trace of the first-order spectra, we centroid the direct image using a 2D Gaussian then apply a position-dependent direct-to-dispersed image offset using the coefficients provided by \citet[Table 1]{Kuntschner2009}.  To find the field-dependent wavelength solution of the observed spectra, we apply the coefficients provided by \citet[Table 5]{Kuntschner2009}.  We model the spectroscopic flat field using the coefficients provided in the file {\tt WFC3.IR.G141.flat.2.fits}.  Additional details on WFC3 calibration can be found in \citet{Berta2012} and references therein.

We extract a 150$\times$150 pixel window centered on the spectrum, of which we use 40 pixels along the spatial direction for our optimal spectral extraction routine and the remainder for background subtraction.  We generate eleven wavelength-dependent light curves spanning 1.10 -- 1.65 {\microns}.  The quality of the data does not warrant using significantly more channels.  We can safely assign one wavelength to each pixel column because the spectrum tilt is only 0.5 -- 0.7\degr from the abscissa \citep{wfc3_dhb} and each binned channel has a width of \sim11 pixels.

\subsection{Light-Curve Systematics}

As previously reported by \citet{Swain2013}, these data do not exhibit a systematic increase in measured fluence between buffer dumps that is seen in other WFC3 exoplanet light curves.  \citet{Swain2013} suggest that this may be connected to the length of time needed for a WFC3 buffer dump and recommend using, at most, the 256$\times$256 subarray mode to minimize systematics.  They also suggest limiting integration times such that the detector receives no more than $\sim$40,000 DN before resetting; this is well above the reported $\sim$33,000 DN (78,000 e\sp{-}) saturation level \citep{wfc3_ihb}.

\citet{Berta2012} plot single-pixel light curves of GJ 1214 as a function of time (see their Figure 4), relative to the first exposure after each buffer dump.  Pixels with fluences in excess of $\sim$40,000 e\sp{-} exhibit a progressively stronger rising exponential ramp, while less-illuminated pixels exhibit no significant trend in time.  Since the maximum pixel fluence in the WASP-12 dataset is $\sim$38,000 e\sp{-}, we should not expect to find any strong systematic behavior.

Our results are also consistent with a WFC3 instrument science report \citep[ISR,][]{Long2013-06} that examines this systematic as it pertains to persistence.  \citet{Long2013-06} examine the count rate as a function of time using six different fluence levels: 28,000, 47,000, 67,000, 87,000, 107,000, and 127,000 e\sp{-}.  Their measurements indicate that a fluence of 28,000 e\sp{-} produces a minuscule change in count rate per unit time that stabilizes within a minute (see their Figure 1, right panel).  Larger fluences produced steeper ramps, indicating greater charge loss for longer periods of time.  These data suggest that future observations should restrict maximum pixel fluence levels to well below 47,000 e\sp{-} ($\sim$19,600 DN).

\subsection{White Light-Curve Fits}
\label{sec:hst-white}

Modeling the band-integrated (white) light curve is an important step to identifying and removing systematics, most of which are wavelength-independent with WFC3.  Additionally, proper modeling is essential to establishing the absolute transit depth when comparing transmission spectra from different instruments at non-overlapping wavelengths. 

\citet{Mandell2013} found evidence for curvature in the out-of-transit data for WASP-12b and WASP-19b.  Both planets are highly irradiated and have relatively short orbital periods; therefore, the effect could be due to thermal emission that varies with orbital phase.  In that case, a sinusoid would be an appropriate function to model this effect.  However, an analysis of the WASP-12b WFC3 emission-spectroscopy data does not reveal the expected complementary trend with negative curvature near peak dayside emission.  This systematic could be the result of long-term thermal fluctuations or short-term thermal settling after telescope repointing.  We elect to explore a variety of ramp models that may produce reasonable fits under these hypotheses.
\citet{Sing2013} report evidence for light-curve fluctuations due to thermal breathing of the telescope as it warms and cools while orbiting the Earth every $\sim$96 minutes.  Such effects are thought to cause small displacements in focus and are often seen in data from other {\em HST} instruments.  To assess the significance of thermal breathing in these data, we choose to model these cyclical variations using a sinusoidal function with a fixed period of 96 minutes.

Table~\ref{tab:hst-fits} summarizes our findings in order of best fit, according to the BIC.  We conclude that inclusion of a thermal-breathing model component offers a statistically significant improvement in the BIC value for most ramp models (except the exponential function).  Although a quadratic ramp with a sinusoidal thermal model achieves the best fit, its $\Delta$BIC value is not significantly different ($\le2.0$) from the exponential ramp models with and without a thermal component.  To investigate this further, we systematically clip more or less data from the fits of the four best models.  In Table~\ref{tab:hst-fixedP}, the parameter $q$ specifies the number of points we have clipped from the start of the observation.  For reference clipping the first orbit is equivalent to a $q$ of 90.  Overall, the exponential ramp model with a thermal component achieves consistent transit depths and the lowest average $\Delta$BIC value.  Therefore, we adopt an uncorrected, band-integrated transit depth of 1.290 {\pm} 0.010\% for our analysis in Section~\ref{sec:hst-nasc}.  This value is consistent with the two other model fits with similar $\Delta$BIC values in Table~\ref{tab:hst-fits} and the relevant transit depths in Table~\ref{tab:hst-fixedP}.  We include the white light curve uncertainty in the absolute transit depth when comparing our results to those from Gemini-N/GMOS and {\em Spitzer}.  The uncertainties on the relative transit depths are valid for comparison within the WFC3 data.

\begin{table}[tb]
\centering
\caption{\label{tab:hst-fits} 
HST/WFC3 White Light-Curve Model Fits}
\begin{tabular}{cccccc}
    \hline
    \hline      
    Label   & Ramp        & Thermal       & rms       & $\Delta$BIC   & Transit Depth \\
            & Model       & Model         & (ppm)     &               & (\%)          \\
    \hline
    a       & Quadratic   & Sinusoidal    & 488       &  0.0          & 1.282 {\pm} 0.014 \\
    b       & Exponential & Sinusoidal    & 488       &  0.2          & 1.290 {\pm} 0.010 \\
    c       & Exponential &  --           & 496       &  2.0          & 1.279 {\pm} 0.008 \\
    d       & Linear      & Sinusoidal    & 497       &  9.3          & 1.337 {\pm} 0.005 \\
    f       & Quadratic   &  --           & 504       & 13.4          & 1.281 {\pm} 0.015 \\
    h       & Linear      &  --           & 513       & 22.5          & 1.334 {\pm} 0.005 \\
    \hline
\end{tabular}
\end{table}

\begin{table}[tb]
\centering
\caption{\label{tab:hst-fixedP} 
Additional White Light-Curve Model Fits With Variable Clipping}
\begin{tabular}{ccccccc}
    \hline
    \hline  
    q:      & 30        & 30            & 60        & 60            & 120       & 120           \\
    Label   & Tr. Depth & $\Delta$BIC   & Tr. Depth & $\Delta$BIC   & Tr. Depth & $\Delta$BIC   \\
            & [\%]      &               & [\%]      &               & [\%]      &               \\
    \hline
    a       & 1.251     &  8.1          & 1.267     &  1.2          & 1.294     &  1.3          \\
    b       & 1.294     &  0.3          & 1.297     &  0.0          & 1.300     &  2.3          \\
    c       & 1.292     &  0.0          & 1.296     &  2.9         & 1.300     &  0.0          \\
    d       & 1.362     & 187.1         & 1.347     & 90.4          & 1.329     &  0.1          \\
    \hline
\end{tabular}
\end{table}

\subsection{Light-Curve Fits (Method 1)}
\label{sec:hst-method1}

As with the GMOS data, we model the light curves using two different techniques then compare the results for consistency.  The first method fits orbits 2 -- 5 of the light curves with a primary-transit model component defined by \citet{Mandel2002}, the same assortment of ramp models listed in Table~\ref{tab:hst-fits}, and a sinusoidal function for the thermal breathing.  We use a quadratic limb-darkening model with fixed parameters, derived from a stellar Kurucz model \citep{Kurucz2004}.  The only wavelength-dependent free parameters are the planet-to-star radius ratio and the absolute flux level.  We do not model the final batch within each orbit of the bluest light curve (1.10 -- 1.15 {\microns}) because its flux is systematically higher than the other batches.  Including these data results in a deeper transit depth at this wavelength.

\begin{figure}[tb]
\centering
\includegraphics[width=1.0\linewidth,clip]{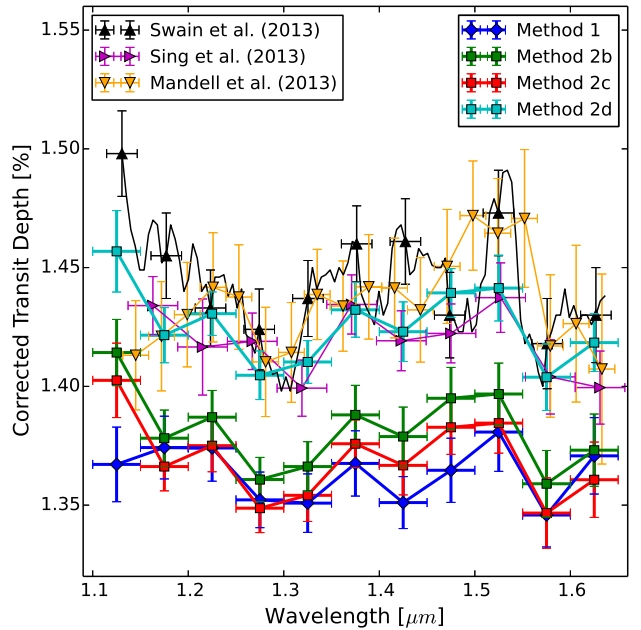}
\caption{\label{fig:hstspec}{
WASP-12b transmission spectrum using WFC3's G141 grism.  Using Method 1, the quadratic (blue diamonds) and exponential ramp models (not shown) achieve comparable best fits with similar transit depths.  For Method 2, the absolute transit depths vary with our choice of ramp model when fitting the white light curve (see Section~\ref{sec:hst-white}).  The green, red, and cyan squares use white light curve transit depths that correspond to labels b, c, and d in Table~\ref{tab:hst-fits}.  All of the relative transit depths using Methods 1 and 2 are in good agreement with previous analyses \citep[][colored triangles]{Swain2013,Sing2013,Mandell2013}; however, only the linear ramp model with a sinusoidal thermal component (Method 2d) agrees on the absolute depths.  These four analyses likely achieve similar transit depths because each one utilizes a linear ramp (either explicitly or implicitly) in its models.
}}
\end{figure}

Similar to the white light curve analysis, including the thermal-breathing model component in our spectroscopic analysis provides a significant improvement in BIC values with all tested ramp models.  The two ramp models that achieve comparable best fits ($\Delta$BIC~$<2.0$) are the quadratic and exponential functions.  Both models favor relatively shallow transit depths compared to previous analyses (see Figure~\ref{fig:hstspec}).  In our spectroscopic analysis, using a linear ramp with a thermal component results in a $\Delta$BIC of $7.0$.

\subsection{Light-Curve Fits (Method 2)}
\label{sec:hst-nasc}

The {\tt Divide-White} method presented here is similar to the one described in Section~\ref{sec:nasc}, with the exception that there are no comparison stars.  Therefore, we skip Step 4 in the synopsis and construct the final light-curve models without the $Z(\lambda,t)$ component.  Again, we use the transit-removed white light curve as our non-analytic model of the wavelength-independent systematics, $Z_{white}(t)$.  With this technique, we find that the spectroscopic light curves exhibit less scatter and can make use of the first orbit.  This second point has significant implications for future observations with WFC3 because this method may reduce the number of orbits needed to make a successful observation if the white light curve transit depth is already known from previous measurements or if only the relative depths are needed.  Light curve fits without data from the first orbit produce similar transit depths (within 1.3$\sigma$) and comparable rms values.

\begin{figure}[tb]
\centering
\includegraphics[width=1.0\linewidth,clip]{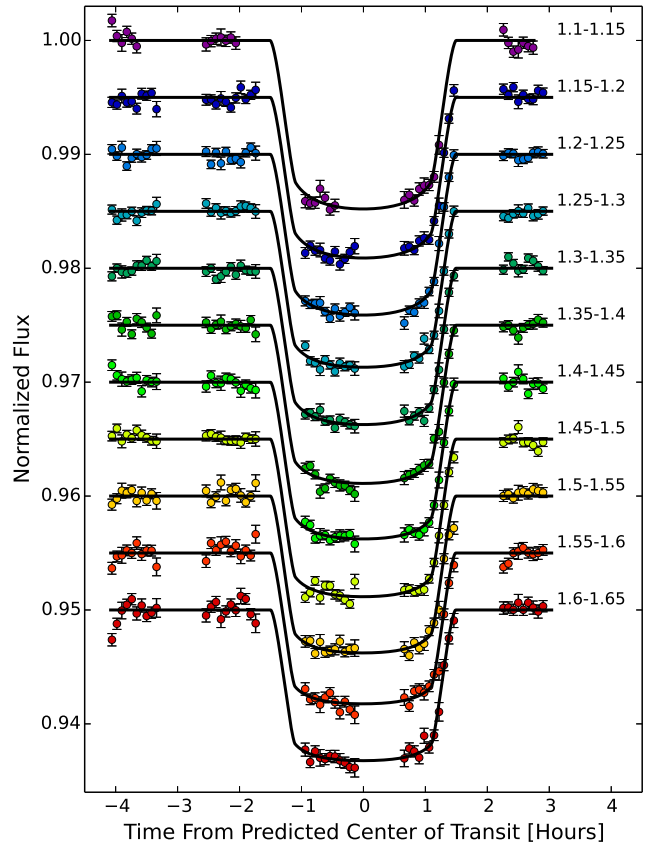}
\caption{\label{fig:hstlc}{
WASP-12b spectroscopic light curves from 2011 May 11 using {\em HST's} WFC3 instrument.  The methods described in Section~\ref{sec:hst-nasc} produced these results, which are binned, normalized to the system flux, and shifted vertically for ease of comparison.  The black lines are best-fit models and the error bars are 1\math{\sigma} uncertainties.  The wavelength range for each of the 11 channels is specified in {\microns}.  We do not model the final batch within each orbit of the bluest channel because its flux is systematically higher than the other batches.  The unbinned residual rms values range from 1190 to 1640 ppm and the average uncertainties are 1.15$\times$ the photon limit.
}}
\end{figure}

We estimate uncertainties with our DE-MCMC algorithm, assuming fixed parameters for $a/R_*$ and $\cos i$ because we are only interested in the relative transit depths.  In agreement with \citet{Swain2013}, correlation plots of rms {\vs}~bin size indicate that there is no significant time-correlated noise in the data and, as such, there is no need to inflate uncertainty estimates \citep{Winn2008b}.  The WFC3 dataset has an insufficient number of points for a robust residual-permutation analysis.  Figure~\ref{fig:hstlc} depicts the normalized WFC3 light curves with best-fit transit models at each wavelength.  The residual rms values range from 1190 to 1640 ppm and the uncertainties range from 1.05 to 1.29$\times$ the photon limit, with an average of 1.15.

In Figure~\ref{fig:hstspec}, we plot the corrected transmission spectrum of WASP-12b and compare our results to those obtained by \citet{Swain2013}, \citet{Sing2013}, and \citet{Mandell2013}.  We find good agreement in the relative transit depths, but the exponential ramp models with and without a thermal component (Methods 2c and 2d) favor significantly lower absolute transit depths, compared to previous analyses, and agree with the absolute depths derived using Method 1 in the previous section.  The most likely reason for this discrepancy is that previous analyses explicitly or implicitly employ a linear ramp (or baseline) in their white and/or spectroscopic light-curve fits.  We recommend that future WFC3 observations acquire additional out-of-transit baseline to more accurately determine the absolute transit depths of WASP-12b.
For the remainder of our analysis, we use the {\tt Divide-White} method derived from the white light curve fit using the exponential ramp model with a thermal component (Method 2c).

%%%%%%%%%%%%%%%%%%%%%%%%%%%%%%%%%%%%%%%%%%%%%%%%%%%%%%%%%%%%%%%%%%%%%%%%%%%%%%%
\section{SPITZER/IRAC OBSERVATIONS AND DATA ANALYSIS}
\label{sec:iracobs}
%%%%%%%%%%%%%%%%%%%%%%%%%%%%%%%%%%%%%%%%%%%%%%%%%%%%%%%%%%%%%%%%%%%%%%%%%%%%%%%

\subsection{Observations and Reduction}

{\em Spitzer's} InfraRed Array Camera \citep[IRAC, ][]{IRAC} observed WASP-12b over its entire orbit at 3.6 and 4.5 {\microns} on 2010 November 17 -- 18 and 2010 December 11 -- 12, respectively (Program number 70060, PI Pavel Machalek).  Each broadband photometric observation acquired $\sim$52,000 frames in sets of 64 (subarray mode).  Exposures within each set are separated by 0.4 seconds and sets are separated by 104-second gaps.

We produce systematics-corrected light curves from {\em Spitzer} Basic Calibrated Data (BCD) files using the Photometry for Orbits, Eclipses, and Transits (POET) pipeline described in detail by \citet{Campo2011} and \citet{Stevenson2011}.  In brief, we flag bad pixels, calculate image centers from a 2D Gaussian fit, and apply interpolated aperture photometry \citep{Harrington2007} over a broad range of aperture sizes.

\subsection{Light-Curve Systematics and Fits}

The steps for modeling the {\em Spitzer} light curves are the same as those described in \citet{Stevenson2011}, with any differences discussed below.  The data do not adequately constrain the stellar limb-darkening coefficients; therefore, we use Kurucz stellar atmospheric models to derive coefficients for a quadratic model \citep{Claret2000}.  The best-fit coefficients are $a_1$,$a_2$ = (0.028575, 0.20868) at 3.6 {\microns} and $a_1$,$a_2$ = (0.028264, 0.17844) at 4.5 {\microns}.

{\em Spitzer} data have well documented systematic effects that our Levenberg-Marquardt minimizer fits simultaneously with the transit parameters.  A linear or quadratic function models the time-dependent systematics and Bilinearly-Interpolated Subpixel Sensitivity (BLISS) mapping \citep{Stevenson2011} models the position-dependent systematics (such as intrapixel variability and pixelation).  We follow the method described by \citet{Stevenson2011} when determining the optimal bin sizes of the BLISS maps.

When estimating uncertainties with our DE-MCMC algorithm, we apply the wavelet technique described by \citet{Carter2009b} to account for correlated noise.  We test all of the available wavelets in the PyWavelets package and find that the Haar wavelet achieves the best fit.  When accounting for correlated noise at 4.5 {\microns}, the transit-depth uncertainty is twice as large as the uncertainty when assuming the light curve only contains white noise.  The residual permutation technique produces smaller uncertainties than the wavelet technique.

\begin{figure}[tb]
\centering
\includegraphics[width=1.0\linewidth,clip]{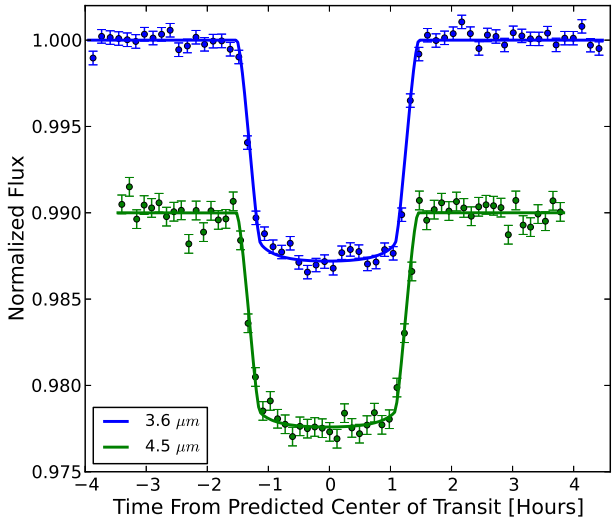}
\caption{\label{fig:spitzerlc}{
WASP-12b photometric light curves using {\em Spitzer's} 3.6 and 4.5 {\micron} channels.  The results are normalized to the system flux and shifted vertically for ease of comparison.  The lines are best-fit models and the error bars are 1\math{\sigma} uncertainties.  
The residual rms values are 0.00617 and 0.00794 at 3.6 and 4.5 {\microns}, respectively.
}}
\end{figure}

In contrast to \citet{Cowan2012}, we do not fit the entire phase curves when determining the transit depths.  This is to prevent unmodeled flux variations in the phase curves [see Figures 4 and 5 from \citet{Cowan2012}] from affecting the measured depths.  We should note that when we do fit the full phase curves, our best-fit models confirm the large ellipsoidal variations reported by \citet{Cowan2012} at 4.5 {\microns}.  At 3.6 and 4.5 {\microns} we fit 15,000 and 13,000 frames, respectively, centered on the transit event.  This allows us to model the local time-dependent flux variations with a linear or quadratic function.  Figure~\ref{fig:spitzerlc} displays the binned, systematics-corrected light curves and their best-fit transit models.  Our uncorrected best-fit transit depth at 3.6 {\microns} is indistinguishable from the value determined by \citet{Cowan2012}; however, our 4.5-{\micron} depth is deeper by 2$\sigma$.  Since we find the same depth as \citet{Cowan2012} when modeling the 4.5 {\micron} phase curves, we attribute this difference to unmodeled flux variations in the phase curves.

%%%%%%%%%%%%%%%%%%%%%%%%%%%%%%%%%%%%%%%%%%%%%%%%%%%%%%%%%%%%%%%%%%%%%%%%%%%%%%%
\section{Dilution Factor Corrections}
\label{sec:dilution}
%%%%%%%%%%%%%%%%%%%%%%%%%%%%%%%%%%%%%%%%%%%%%%%%%%%%%%%%%%%%%%%%%%%%%%%%%%%%%%%

\subsection{Correcting for WASP-12's Stellar Companion}
\label{sec:dilution1}

WASP-12 is a hierarchical triple star system in which the secondary and tertiary companions (WASP-12BC) are $\sim$1{\arcsec} in separation from the primary \citep{Bechter2013}.  Because of the companions' proximity, we cannot mask it's contribution to the spectra or the resulting light curves in any of the data sets, so we correct the measured transit depths using stellar atmospheric models to compute a wavelength-dependent dilution factor.  We assume both companions have the same stellar type of M0 -- M1 \citep{Crossfield2012}.

To measure the dilution factor ($\alpha_{Comp} = F_{Comp}/F_{W12}$), we use the WFC3 photometric image (acquired using the F132N filter, see Figure~\ref{fig:wa012b-hst-phot}) to centroid both WASP-12 and the companion star(s) with 2D Gaussians, then perform aperture photometry with radii of 3.0 and 2.0 pixels, respectively.  Companion apertures of $\leq$2.0 pixels exhibit negligible contamination from WASP-12.  Dilution factor values are consistent (within 1$\sigma$) for WASP-12 aperture sizes in the range of 2.0 -- 4.0 pixels.  We correct for the unaccounted flux outside the apertures by dividing each measured flux by the theoretical aperture fraction, as determined by point spread functions (PSFs) computed using the Tiny Tim software package \citep{Krist2011} at each object's position on the detector.  We determine the dilution factor at 1.32 {\microns} to be 0.0692 {\pm} 0.0015.

\begin{figure}[tb]
\centering
\includegraphics[width=1.0\linewidth,clip]{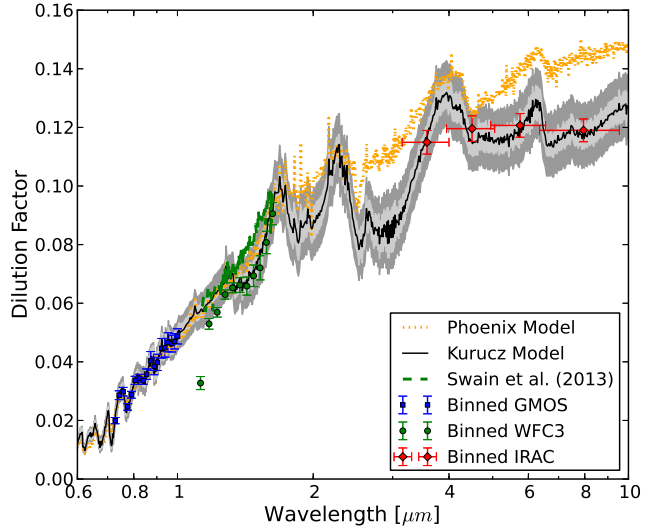}
\caption{\label{fig:dilfactor}{
Wavelength-dependent dilution factors due to WASP-12's stellar companion.  WASP-12 has a spectral type of G0 and its companion is M0 -- M1.  The black line depicts the calculated dilution factors using a Kurucz model with stellar parameters from \citet{Hebb2009}.  The dark gray region represents the absolute uncertainties and the light gray region represents the uncertainties relative to the measured dilution factor at 1.32 {\microns}.  The colored points represent binned dilution factors with uncertainties for each channel of each instrument discussed in this paper.  Section~\ref{sec:dilution1} describes why the employed WFC3 dilution factors drop off near 1.1 {\microns}.  For comparison, the dashed green line depicts the dilution factors derived by \citet{Swain2013} using PSF fitting and the dotted orange line displays the calculated dilution factors using a PHOENIX model from the BT-Settl library \citep{Allard2011} with $T\sb{W12}=6300$ K and $T\sb{Comp}=3700$ K.  The Kurucz and PHOENIX models agree (to within the uncertainties) below \sim1.7 {\microns}; however, there are clear discrepancies at the longer {\em Spitzer} wavelengths.  Because the uncertainties in the {\em Spitzer} points are larger than these discrepancies, our choice to use the Kurucz stellar model is moot.
}}
\end{figure}

\begin{table*}[tb]
\centering
\caption{\label{tab:depths} 
WASP-12\MakeTextLowercase{b} Transit Depths and Dilution Factors}
\begin{tabular}{cr@{\,{\pm}\,}lr@{\,{\pm}\,}lr@{\,{\pm}\,}lr@{\,{\pm}\,}l}
    \hline
    \hline      
    Bandpass            & \mctc{Meas. Depth}    & \mctc{Companion Star}     & \mctc{Planet Nightside}   & \mctc{Corr. Depth\tablenotemark{a}} \\
    ({\microns})        & \mctc{(\%)}           & \mctc{Dilution Factor}    & \mctc{Dilution Factor}    & \mctc{(\%)}               \\
    \hline
    0.720 -- 0.735      & 1.381 & 0.021         & 0.0199 & 0.0011           & 0.00039 & 0.00004         & 1.409 & 0.021             \\
    0.735 -- 0.750      & 1.399 & 0.011         & 0.0286 & 0.0015           & 0.00091 & 0.00008         & 1.441 & 0.012             \\
    0.750 -- 0.765      & 1.381 & 0.006         & 0.0298 & 0.0014           & 0.00097 & 0.00008         & 1.423 & 0.006             \\
    0.768 -- 0.783      & 1.383 & 0.006         & 0.0246 & 0.0011           & 0.00057 & 0.00005         & 1.417 & 0.007             \\
    0.783 -- 0.798      & 1.384 & 0.008         & 0.0288 & 0.0012           & 0.00082 & 0.00007         & 1.425 & 0.008             \\
    0.798 -- 0.813      & 1.392 & 0.006         & 0.0337 & 0.0014           & 0.00117 & 0.00009         & 1.441 & 0.006             \\
    0.817 -- 0.832      & 1.384 & 0.005         & 0.0344 & 0.0013           & 0.00122 & 0.00010         & 1.434 & 0.006             \\
    0.832 -- 0.847      & 1.364 & 0.008         & 0.0335 & 0.0012           & 0.00114 & 0.00009         & 1.411 & 0.009             \\
    0.847 -- 0.862      & 1.362 & 0.006         & 0.0358 & 0.0018           & 0.00111 & 0.00009         & 1.412 & 0.007             \\
    0.865 -- 0.880      & 1.361 & 0.008         & 0.0403 & 0.0032           & 0.00126 & 0.00010         & 1.418 & 0.009             \\
    0.880 -- 0.895      & 1.350 & 0.007         & 0.0385 & 0.0029           & 0.00107 & 0.00009         & 1.403 & 0.008             \\
    0.895 -- 0.910      & 1.349 & 0.007         & 0.0398 & 0.0027           & 0.00112 & 0.00009         & 1.404 & 0.008             \\
    0.915 -- 0.930      & 1.374 & 0.006         & 0.0446 & 0.0033           & 0.00152 & 0.00012         & 1.437 & 0.008             \\
    0.930 -- 0.945      & 1.366 & 0.008         & 0.0448 & 0.0032           & 0.00152 & 0.00013         & 1.429 & 0.009             \\
    0.945 -- 0.960      & 1.357 & 0.012         & 0.0467 & 0.0032           & 0.00165 & 0.00013         & 1.422 & 0.014             \\
    0.963 -- 0.978      & 1.408 & 0.013         & 0.0462 & 0.0028           & 0.00166 & 0.00013         & 1.476 & 0.014             \\
    0.978 -- 0.993      & 1.387 & 0.010         & 0.0470 & 0.0026           & 0.00166 & 0.00013         & 1.454 & 0.011             \\
    0.993 -- 1.008      & 1.356 & 0.014         & 0.0487 & 0.0025           & 0.00173 & 0.00013         & 1.424 & 0.015             \\
    \hline
    1.100 -- 1.150      & 1.375 & 0.013         & 0.0328 & 0.0022           & 0.00241 & 0.00016         & 1.414 & 0.013             \\
    1.150 -- 1.200      & 1.315 & 0.010         & 0.0530 & 0.0018           & 0.00262 & 0.00016         & 1.378 & 0.011             \\
    1.200 -- 1.250      & 1.318 & 0.010         & 0.0569 & 0.0016           & 0.00286 & 0.00017         & 1.387 & 0.010             \\
    1.250 -- 1.300      & 1.286 & 0.008         & 0.0629 & 0.0015           & 0.00301 & 0.00017         & 1.361 & 0.009             \\
    1.300 -- 1.350      & 1.288 & 0.009         & 0.0653 & 0.0017           & 0.00327 & 0.00017         & 1.366 & 0.010             \\
    1.350 -- 1.400      & 1.307 & 0.009         & 0.0662 & 0.0025           & 0.00342 & 0.00016         & 1.388 & 0.011             \\
    1.400 -- 1.450      & 1.298 & 0.009         & 0.0658 & 0.0031           & 0.00366 & 0.00017         & 1.379 & 0.011             \\
    1.450 -- 1.500      & 1.309 & 0.010         & 0.0694 & 0.0037           & 0.00398 & 0.00017         & 1.395 & 0.012             \\
    1.500 -- 1.550      & 1.307 & 0.011         & 0.0721 & 0.0041           & 0.00450 & 0.00018         & 1.397 & 0.012             \\
    1.550 -- 1.600      & 1.261 & 0.013         & 0.0807 & 0.0039           & 0.00508 & 0.00020         & 1.359 & 0.013             \\
    1.600 -- 1.650      & 1.262 & 0.012         & 0.0905 & 0.0038           & 0.00568 & 0.00021         & 1.373 & 0.013             \\
    \hline
    3.6\tablenotemark{b}& 1.232 & 0.018         & 0.1149 & 0.0039           & 0.00651 & 0.00024         & 1.341 & 0.020             \\
    4.5\tablenotemark{c}& 1.199 & 0.028         & 0.1196 & 0.0042           & 0.00614 & 0.00024         & 1.306 & 0.031             \\
    5.8                 & \mctc{--}             & 0.1207 & 0.0039           & 0.00653 & 0.00025         & \mctc{--}                 \\
    8.0                 & \mctc{--}             & 0.1190 & 0.0038           & 0.00633 & 0.00023         & \mctc{--}                 \\
    \hline
\end{tabular}
\tablenotetext{1}{Since we are primarily concerned with the relative transit depths, we fix $\cos i$ and $a/R_*$ to 0.164 and 2.908, respectively.}
\tablenotetext{2}{Dilution factor is multiplied by $g(2.25,3.6)=0.7116$, the calculated companion fraction inside an aperture of 2.25 pixels at 3.6 {\microns}.}
\tablenotetext{3}{Dilution factor is multiplied by $g(2.25,4.5)=0.6931$, the calculated companion fraction inside an aperture of 2.25 pixels at 4.5 {\microns}.}
\end{table*}

To assess the dilution factor at other wavelengths (see Figure~\ref{fig:dilfactor} and Table~\ref{tab:depths}), we first use Kurucz stellar atmospheric models of WASP-12 ($K_{W12}$) and its companions ($K_{Comp}$) at 1.32 {\microns} to calculate the wavelength-independent geometric ratio:
\begin{equation}
\label{eqn:georatio}
f^2 \equiv \left(\frac{R_{Comp}}{R_{W12}}\frac{d_{W12}}{d_{Comp}}\right)^2 = \alpha_{Comp}(\lambda)\frac{K_{W12}(\lambda)}{K_{Comp}(\lambda)}.
\end{equation}
\noindent Here, $R$ is the stellar radius and $d$ is the distance from the observer.  The geometric ratio adjusts the theoretical flux values to account for the positions and sizes of the stellar objects.  Given that $d_{W12} = d_{Comp}$ and assuming WASP-12BC are the same spectral type, we find their radii to be 0.56 {\pm} 0.03~R\sb{$\odot$}.

Once we have the geometric ratio, we apply Equation \ref{eqn:georatio} to calculate the dilution factor at other wavelengths.  We estimate dilution factor uncertainties through bootstrapping, wherein we generate 5,000 atmospheric models with a distribution of stellar temperatures ($T\sb{W12}=6300{\pm}150$ K, $T\sb{Comp}=3660{\pm}70$ K), then measure the distribution of dilution factors at each wavelength, using the 1.32-{\micron} value as an anchor.  The models closely agree with $K$-band measurements of the dilution factor from \citet{Crossfield2012}; however, they are higher than the $i^{\prime}$- and $z^{\prime}$-band measurements from \citet{Bergfors2013} by 1.4 and 2.3$\sigma$, respectively.  The dilution factors derived by \citet{Swain2013} using PSF fitting are systematically higher than both the Phoenix and Kurucz models for wavelengths greater than 1.3 {\microns} (see Figure~\ref{fig:dilfactor}).

The {\em HST/WFC3} data require an additional step during the correction process because WASP-12BC's spectrum is not spectroscopically aligned on the detector with that of WASP-12.  Using the measured separation along the dispersion direction from the frame depicted in Figure~\ref{fig:wa012b-hst-phot}, we estimate that contamination from the stellar companion is shifted redward by 0.0324 {\microns}.  To calculate the companion star dilution factor, we multiply the WASP-12 and red-shifted companion Kurucz spectra by the WFC3 G141 transmission filter to properly weight their respective contributions.  The effect of these additional steps is most readily seen in the 1.1 -- 1.15 {\micron} channel in Table~\ref{tab:depths} where there is noticeably less contamination from the companion star.

\subsection{Correcting for WASP-12b's Nightside Emission}

In addition to correcting for the presence of a stellar companion, we also consider the wavelength-dependent effects of emission from WASP-12b's nightside on the measured transit depths \citep{Kipping2010}.  The magnitude of this correction is typically negligible; however, the effect is noticeable for a highly irradiated object such as WASP-12b over a broad spectral range.  To begin, we generate a Kurucz atmospheric model using a hemispherically-averaged effective nightside temperature of 983 {\pm} 200 K \citep{Cowan2012} and a planetary surface gravity ($\log g_{p}$) of 2.99 {\pm} 0.03 \citep{Hebb2009}.  We then apply Equation \ref{eqn:georatio} with $f = R_p/R_*$ from Table~\ref{tab:transitparams} to estimate the planet nightside dilution factor, $\alpha_{p}(\lambda)$.  As in Section~\ref{sec:dilution1}, we estimate dilution factor uncertainties through bootstrapping.  We list the wavelength-dependent planet nightside dilution factors with uncertainties in Table~\ref{tab:depths}.

\subsection{Corrected Transit Depths}

We determine the corrected transit depths by:
\begin{equation}
\label{eqn:trdepth}
\delta\sb{Corr}(\lambda) = \left [1+g(\beta,\lambda)\alpha_{Comp}(\lambda)+\alpha_{p}(\lambda)\right]\delta\sb{Meas}(\lambda),
\end{equation}
\noindent where $g(\beta,\lambda)$ are the wavelength-dependent companion flux fractions inside a photometric aperture of size $\beta$ and $\delta\sb{Meas}(\lambda)$ are the measured (uncorrected) transit depths.  To estimate $g(\beta,\lambda)$, we use the 5$\times$oversampled {\em Spitzer} point response functions (PRFs) calculated at pixel position (25,25).  We set $g(\beta,\lambda)$ to unity for spectroscopic analyses.  Table~\ref{tab:depths} presents the corrected transit-depth values and uncertainties for observations from all three instruments discussed in the previous sections.

\begin{figure}[tb]
\centering
\includegraphics[width=1.0\linewidth,clip]{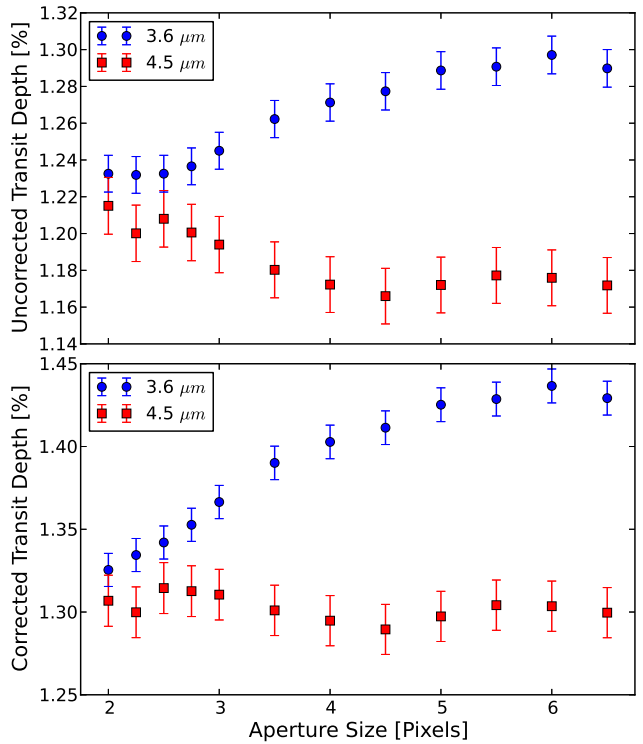}
\caption{\label{fig:apsize}{
Measured {\em Spitzer} transit depths for various photometry aperture sizes.  We plot uncorrected (top panel) and WASP-12BC-corrected (bottom panel) transit depths at 3.6 and 4.5 {\microns} (blue circles and red squares, respectively).  We also account for $g(\beta,\lambda)$, the fraction of WASP-12BC flux that falls inside each photometric aperture.  As expected, the measured 4.5-{\micron} uncorrected transit depths decrease with increasing aperture size as the aperture captures additional flux from WASP-12BC.  This is confirmed by the relatively flat, aperture-independent 4.5-{\micron} transit depths in the corrected panel. Conversely, the measured 3.6-{\micron} uncorrected transit depths increase with increasing aperture size, which would require a physically impossible negative flux from WASP-12BC.  This effect may be due to an observed increase in correlated noise with increasing aperture size.  For comparison, the IRAC plate scale is \sim1.2~\arcsec/pixel and the full width at half maximum of the PRF is \sim1.7\arcsec.
}}
\end{figure}

To avoid introducing additional uncertainty through correcting the dilution factor for the fraction of flux that falls inside of {\em Spitzer's} aperture, we would ideally use sufficiently large apertures that capture all of the flux from both WASP-12 and its companion \citep{Crossfield2012}.  However, we find that the measured transit depths at 3.6~{\microns} do not decrease as expected with larger aperture sizes (see Figure~\ref{fig:apsize}).  Instead, they increase out to \sim5-pixel apertures before plateauing.  This effect suggests that other systematics have a more significant impact on the measured transit depth than the dilution by WASP-12BC as a function of aperture size.  One possible cause is an observed increase in correlated noise within the 3.6-{\micron} transit as we increase the aperture size.  Alternatively, the effect may be due to ineffective modeling of the position-dependent systematics.  As a result of the aperture-dependent transit depths at 3.6~{\microns}, we choose to use the best aperture size (2.25 pixels in both channels) according to the lowest standard deviation of the normalized residuals (SDNR) then apply a correction for the fraction of light from WASP-12BC that falls inside of the aperture (0.7116 and 0.6931 at 3.6 and 4.5 {\microns}, respectively).

%%%%%%%%%%%%%%%%%%%%%%%%%%%%%%%%%%%%%%%%%%%%%%%%%%%%%%%%%%%%%%%%%%%%%%%%%%%%%%%
\section{ATMOSPHERIC MODELS AND DISCUSSION}
\label{sec:results}
%%%%%%%%%%%%%%%%%%%%%%%%%%%%%%%%%%%%%%%%%%%%%%%%%%%%%%%%%%%%%%%%%%%%%%%%%%%%%%%

We apply the atmospheric modeling and retrieval technique described by \citet{Madhu2012} to place constraints on the properties of WASP-12b's atmosphere.  Under the conditions of local thermodynamic equilibrium, hydrostatic equilibrium, and global energy balance, we compute model spectra using 1-D line-by-line radiative transfer in a plane-parallel atmosphere.  This approach makes no assumptions about the atmospheric chemical abundances or layer-by-layer radiative equilibrium.  The model atmospheres include molecular absorption due to all the major molecules expected to be dominant in O-rich and C-rich atmospheres of hot Jupiters \citep[e.g. ][]{Madhu2011-Crich,Kopparapu2012,Madhu2012, Moses2013}, namely, H$_2$O, CO, CH$_4$, CO$_2$, HCN, C$_2$H$_2$, TiO, VO, Na, K, and H$_2$-H$_2$ collision-induced absorption.  The sources for the molecular line-lists are discussed in \citet{Madhu2012}.  We also include line absorption due to TiH based on opacities from \citet{Burrows2005TiH} and the cross-sections computed from \citet{Hill2012}.  For comparison, we also compute simpler, physically-implausible atmospheric models containing pure H\sb{2} and H\sb{2} plus K.  The thermal profile is defined by six points at different pressures that are constrained from previous analyses of available thermal-emission data \citep{Madhu2011-wasp12b}.  This is because the thermal profile is poorly constrained with transmission spectroscopy.

\subsection{Cloud-Free Atmospheric Models}
\label{sec:nominal}

\begin{table}[tb]
\centering
\caption{\label{tab:chi2} 
$\chi^2$ Values Fitting Select Groups of Data.}
\begin{tabular}{cccc}
    \hline
    \hline      
    Model       & GMOS Only & WFC3 Only & All Data  \\
    \hline
    O-rich      &  83       &  18       & 143       \\
    C-rich      & 106       &  33       & 188       \\
    H\sb{2}+K   &  87       &  21       & 169       \\
    H\sb{2}     &  55       &  22       & 224       \\
    \hline
\end{tabular}
\end{table}

\begin{figure}[tb]
\centering
\includegraphics[width=1.0\linewidth,clip]{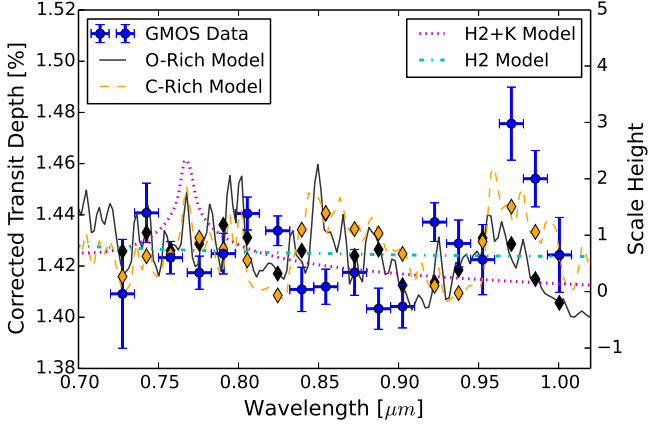}
\caption{\label{fig:corrspec-gmos}{
WASP-12b corrected GMOS transmission spectrum with atmospheric models.  The blue GMOS data points have error bars that depict the wavelength bin size and transit depth 1$\sigma$ uncertainties.  The dotted magenta and dashed-dotted cyan lines depict a H\sb{2} atmospheric model with and without a broad potassium (K) resonance doublet at \sim0.77~{\microns}.  The O-rich (solid black line) and C-rich (dashed orange line) models contain metal oxides and metal hydrides, respectively.  For ease of comparison, the colored diamonds represent bandpass-integrated models.  The GMOS results agree with the z'-band corrected transit depth (1.44 {\pm} 0.03\%, Leslie Hebb, private communication, 2013) from the discovery paper.  The atmospheric models are vertically offset from Figure~\ref{fig:corrspec} to fit only the GMOS data.
}}
\end{figure}

\begin{figure}[tb]
\centering
\includegraphics[width=1.0\linewidth,clip]{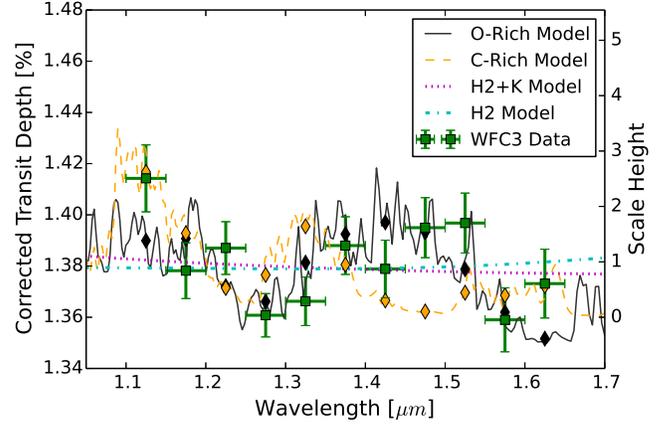}
\caption{\label{fig:corrspec-hst}{
WASP-12b corrected WFC3 transmission spectrum with atmospheric models.  Error bar and atmospheric model definitions are the same as in Figure~\ref{fig:corrspec-gmos}.  The O-rich model contains H\sb{2}O, whereas the C-rich model contains CH\sb{4} and HCN.  For ease of comparison, the colored diamonds represent bandpass-integrated models.  The atmospheric models are vertically offset from Figure~\ref{fig:corrspec} to fit only the WFC3 data.
}}
\end{figure}

\begin{figure}[tb]
\centering
\includegraphics[width=1.0\linewidth,clip]{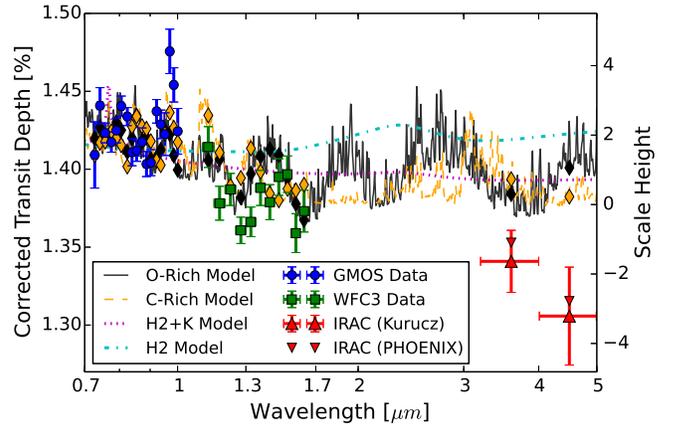}
\caption{\label{fig:corrspec}{
WASP-12b corrected transmission spectrum with cloud-free atmospheric models.  The data include Gemini-N/GMOS observations (blue circles) in the red optical, {\em HST/WFC3} observations (green squares) in the NIR, and {\em Spitzer/IRAC} observations (red triangles, Kurucz model) from 3 -- 5 {\microns}.  A second set of red triangles without uncertainties utilize a PHOENIX stellar model to correct for the companion star (see Section~\ref{sec:dilution}); results from both stellar models are consistent.  The solid black line depicts an atmospheric model with a solar C/O and the dashed orange line corresponds to a planetary atmosphere with a C/O = 1.  In comparing the data to the bandpass-integrated models (colored diamonds), we see that the available data at their current precision do not clearly favor one model over the other.  For reference, we also include H\sb{2} atmospheric models with and without the broad K resonance doublet at \sim0.77~{\microns}.
}}
\end{figure}

In this section, we consider cloud-free atmospheric models.  When fitting the absolute scale of the atmospheric models to the data, we take into account the uncertainty in the absolute transit depths by allowing each set of GMOS and WFC3 depths to shift up or down while applying priors according to the uncertainties on the white light curve depths.  We find that neither realistic atmospheric model significantly outperforms the other when fitting the data.  This can be seen in Table~\ref{tab:chi2} and in Figures \ref{fig:corrspec-gmos} -- \ref{fig:corrspec}, where we fit only the GMOS data, only the WFC3 data, and all of the data.  The C-rich atmospheric model (C/O = 1) contains numerous metal hydride features (due to TiH) in the red optical and broader CH\sb{4} and HCN features in the NIR.  This model fits the 31 data points with a $\chi^2$ value of 188 and 21 degrees of freedom (DoF).  The O-rich model (C/O = 0.5) favors metal oxides, such as TiO and VO, in the red optical and a broad H\sb{2}O feature in the NIR.  This model produces a slightly more favorable $\chi^2$ of 143 with 24 DoF.  We caution that the apparent H\sb{2}O feature in the O-rich model may be enhanced by the dilution correction from WASP-12BC (see Figure~\ref{fig:dilfactor}).  Given the quality of these fits, estimating molecular abundances with a full atmospheric retrieval is unwarranted as the results would have a high degree of inaccuracy.  We may be able to improve the fits to the GMOS data by arbitrarily adding additional metal-hydride or metal-oxide opacity sources; however, this will not help us to distinguish between the two prevailing atmospheric models.  We do not detect the strong potassium resonance doublet at $\sim$0.77 {\microns} because it is likely being masked by the many metal oxide or hydride lines in that part of the spectrum (see Figure~\ref{fig:corrspec-gmos}).  We determine that the corrected WASP-12b transmission spectrum rules out a cloud-free, H\sb{2} atmosphere with no additional opacity sources ($\chi^2 = 224$). The addition of K (2\tttt{-7} relative abundance) to a H\sb{2} atmosphere produces a slightly better fit ($\chi^2 = 169$).  Although both fits are comparable to the physically plausible O- and C-rich models, the realistic models may be improved by adding additional sources of opacity or enhanced scattering.

\begin{figure*}[tb]
\centering
\includegraphics[width=1.0\linewidth,clip]{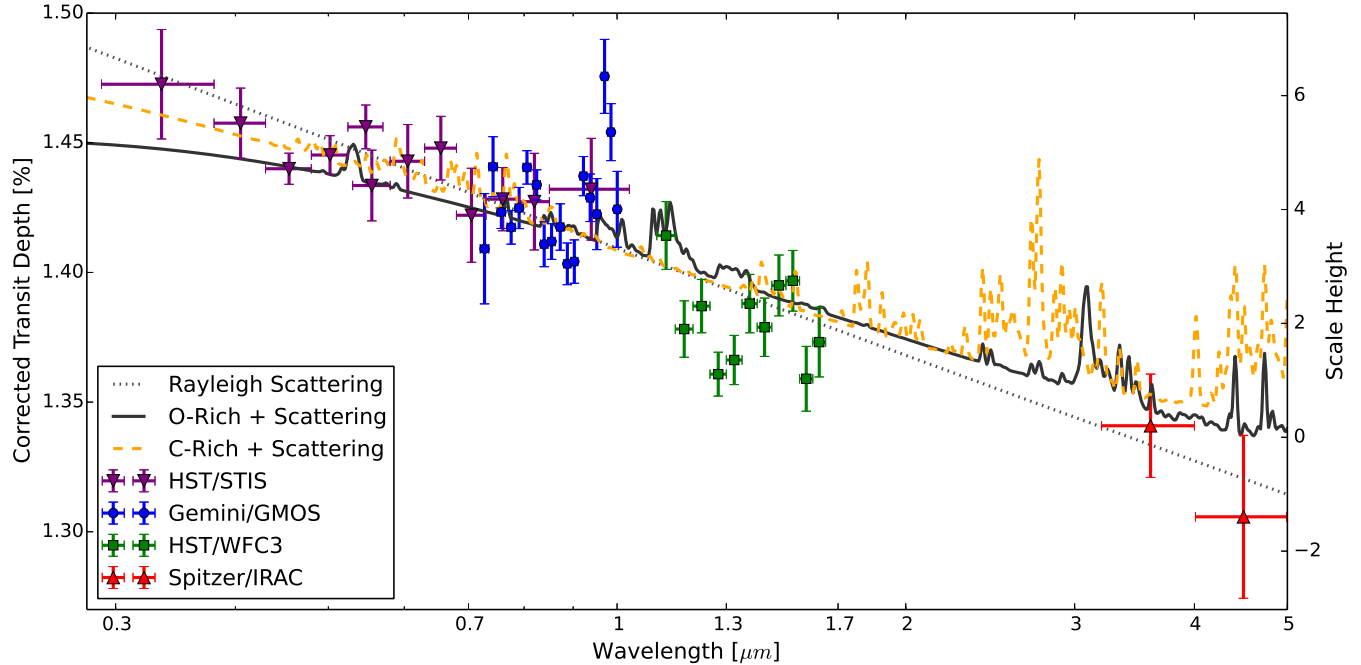}
\caption{\label{fig:finalspec}{
WASP-12b corrected transmission spectrum with best-fit enhanced scattering atmospheric models.  In addition to the data in Figure~\ref{fig:corrspec}, we include published results from \citet[][inverted purple triangles]{Sing2013}.  The O- and C-rich models achieve comparable fits, but require relatively low reference pressures ($\leq 5$~mbar).  A linear fit representing an idealized scattering model (dotted line) achieves the best fit with a $\chi^2_{\nu} = 3.0$.
}}
\end{figure*}

After correcting for contribution from the planet nightside and contamination from WASP-12BC, and including those uncertainties in our corrected transit depth uncertainties, both O- and C-rich cloud-free models fit many of the Gemini-N/GMOS and {\em HST/WFC3} transit depths to within 2$\sigma$.  However, the models consistently over-predict the {\em Spitzer/IRAC} transit depths.  This can be seen in Figure~\ref{fig:corrspec}, which depicts a decreasing trend in transit depth with increasing wavelength.  If real, the transmission spectrum spans approximately eight scale heights over these wavelengths.  This trend may be the result of enhanced scattering from atmospheric hazes or aerosols.  In Section~\ref{sec:scattering}, we explore different scattering models for WASP-12b in detail and compare our findings to those of \citet{Sing2013}.

We also consider two plausible scenarios in which the observed trend in Figure~\ref{fig:corrspec} is not real.  First, this trend may be the result of residual contamination from WASP-12BC.  An observed spectrum of the companions would help to confirm this theory.  Second, since the observations considered in this paper were not all acquired simultaneously, it is conceivable that stellar variability may introduce a vertical offset between data sets.  Variability has been noted as a potential source of disagreement in other exoplanet systems \citep{Knutson2011,Ballerini2012}.  Follow-up observations with GMOS and WFC3 would place constraints on stellar variability.

\subsection{Enhanced Scattering}
\label{sec:scattering}

In this section, we consider several realistic and toy models with enhanced scattering to fit the measured WASP-12b transmission spectrum.  For completeness, this analysis includes the HST/STIS results presented by \citet{Sing2013}.

First, we fit the slope of the transmission spectrum assuming an arbitrary atmospheric opacity source with an effective extinction cross section, $\sigma = \sigma_0 (\lambda/\lambda_0)^{\alpha}$ \citep{Lecavelier2008}.  The slope is related to $\alpha$ and $T$ as follows:

\begin{equation}
\alpha H = \alpha \frac{kT}{\mu g} = \frac{d R_p}{d \ln \lambda}
\end{equation} 

\noindent Assuming a mean molecular weight, $\mu = 2.3$ g/mol, and a planetary surface gravity, $g = 2.99$ dex \citep{Hebb2009}, we estimate $\alpha T = -7460 {\pm} 520$.  This slope is significantly steeper than that estimated by \citet[][$\alpha T = -3528 {\pm} 660$]{Sing2013} due to the lower absolute transit depths favored in our {\em HST} and {\em Spitzer} analyses.  As a result, we find a more realistic estimate of the planet's terminator temperature ($T_{term} = 1870 {\pm} 130$ K) when assuming the slope can be described by Rayleigh scattering ($\alpha = -4$, see Figure~\ref{fig:finalspec}).  Conversely, if we assume a temperature of 2100 K, this implies $\alpha = -3.55 {\pm} 0.25$.  In either case, we achieve a $\chi^2$ of 123 with 41 DoF (see Table~\ref{tab:chi2-scattering}).

\begin{table}[tb]
\centering
\caption{\label{tab:chi2-scattering} 
$\chi^2$ Values Considering Select Scattering Models.}
\begin{tabular}{cccc}
    \hline
    \hline      
    Model               & $\chi^2$  & k, DoF    & $\Delta$BIC   \\
    \hline
    Rayleigh Scattering & 123       &  2, 41    &  0            \\
    O-Rich + Scattering & 137       &  7, 36    & 33            \\
    C-Rich + Scattering & 135       & 10, 33    & 42            \\
    \hline
\end{tabular}
\end{table}

Second, we explore adding scattering from an arbitrary atmospheric opacity source to our O- and C-rich models from Section~\ref{sec:nominal}.  Using $\alpha = -3$, a cross section $0.1\times$ that of H\sb{2}, and $T_{term} = 2200$ K, our best-fit C-rich model achieves a $\chi^2$ values of 135 with 33 DoF and our best-fit O-rich model achieves a $\chi^2$ values of 137 with 36 DoF (see Figure~\ref{fig:finalspec}).  However, both models rely on relatively low reference pressures of $\leq 5$~mbar and a high terminator temperature.  The reference pressure is the level at which we set the atmospheric model altitude equal to the measured planet radius \citep{Hebb2009}.  For comparison, we typically set the reference pressure to $\sim$1~bar.  However, the uncertainty on the measured planet radius allows for a large range of plausible reference pressures.

\begin{figure}[tb]
\centering
\includegraphics[width=1.0\linewidth,clip]{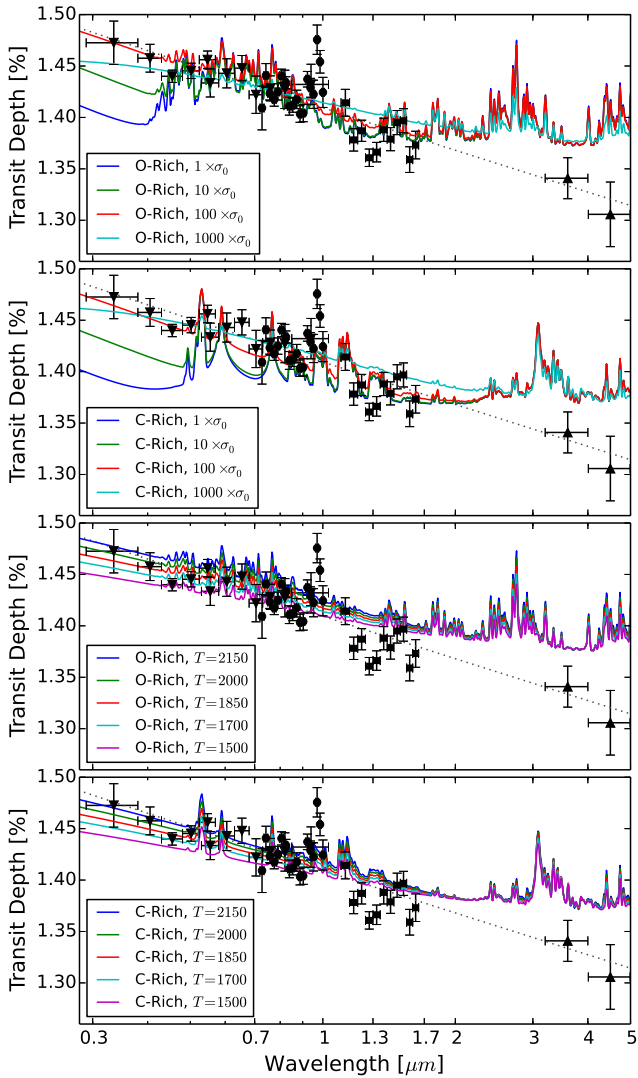}
\caption{\label{fig:scattering}{
WASP-12b corrected transmission spectrum with additional enhanced scattering atmospheric models.
The top two panels explore the effects of increasing the effective H\sb{2} cross section for both O- and C-rich models (colored lines), assuming $\alpha = -4$.  The O-rich models require a cross section at least $100\times\sigma_0$ to obscure the potential presence of TiO, as seen in the STIS data by the absence of a blue edge in the TiO bandhead near 0.43 {\microns}.  The third and fourth panels illustrate that $0.04\times$ and $0.1\times$ H\sb{2} effective cross sections with a power law index $\alpha = -3$ can also produce reasonable fits to the non-{\em Spitzer} data over a range of acceptable terminator temperatures.  More precise data are needed over a broad wavelength range to distinguish between the assortment of models presented here.
}}
\end{figure}

Finally, using a more stringent, 20~mbar upper limit on the reference pressure,we examine additional scattering models within our parameter space (see Figure~\ref{fig:scattering}).  Assuming a fixed $\alpha$ of -4, we test models with a range of effective cross sections and suitable terminator temperatures.  We find that O- and C-rich models with at least $100\times$ H\sb{2} cross section can adequately fit the STIS, GMOS, and WFC3 data.  We also explore scattering models with other power law indices and effective cross sections.  Figure~\ref{fig:scattering} provides illustrative examples of the most reasonable fits.  However, in all instances, the $\chi^2$ values are higher than those reported in Table~\ref{tab:chi2-scattering}.  This is because none of the models can explain the relatively low transit depths in the 3.6 and 4.5 {\micron} bandpasses.

The identity of the scattering source remains a mystery.  Corundum and silicate condensates may have difficulty forming given WASP-12b's high terminator temperature.  Furthermore, strong vertical mixing is required to keep these heavy molecules suspended at high altitudes \citep{Spiegel2009}.  One potentially viable explanation for the observed transmission spectrum is the formation of hydrocarbon soot-like particles \citep{Morley2013}.  More work is needed to fully explore these options.

%%%%%%%%%%%%%%%%%%%%%%%%%%%%%%%%%%%%%%%%%%%%%%%%%%%%%%%%%%%%%%%%%%%%%%%%%%%%%%%
\section{CONCLUSIONS}
\label{sec:concl}
%%%%%%%%%%%%%%%%%%%%%%%%%%%%%%%%%%%%%%%%%%%%%%%%%%%%%%%%%%%%%%%%%%%%%%%%%%%%%%%

In this paper, we present new, high-precision transmission spectra of the highly irradiated exoplanet WASP-12b using the technique of wide-slit, multi-object spectroscopy in the red optical.  Both Gemini-N nights produce comparable results in terms of the measured transit depths, thus providing confidence in our analysis and validating the consistency and reliability of the Gemini-N/GMOS instrument.  We model the spectroscopic light curves using two different techniques, one of which is new and both of which produce similar best-fit results.  We supply step-by-step instructions for this newly developed technique, which models the spectroscopic light curves without making any prior assumptions about the form of the systematics and produces smaller residual rms values than the other technique.

We also present a reanalysis of previously published WASP-12b transit data from {\em HST} and {\em Spitzer}.  We correct all of the measured transit depths by wavelength-dependent dilution factors to account for the nearby stellar companions to WASP-12 and for the emission from WASP-12b's nightside.  For the {\em HST/WFC3} data, we find that the absolute transit depths are systematic-model dependent.  Applying different combinations of ramp and thermal models produces best-fit solutions with comparable BIC values but with white light curve transit depths that vary by as much as 0.055\%.  We recommend that future WFC3 observations acquire additional out-of-transit baseline to more accurately establish the absolute transit depths of WASP-12b.

We model the corrected transit depths using a retrieval-based technique that considers sources of opacity for both O- and C-rich atmospheres.  The GMOS and WFC3 data sets may be explained by an O-rich atmosphere in which TiO, VO, and H\sb{2}O dominate in the red optical and NIR.  A C-rich atmosphere in which TiH, CH\sb{4}, and HCN dominate the spectrum may equally explain the two data sets.  However, the inclusion of STIS data in the optical and IRAC data in the NIR indicate the presence of a slope, possibly due to scattering.  The available data exclude a featureless, pure H\sb{2} atmosphere.  Our best-fit model is a linear trend that contains no molecular absorption features.  Physically-motivated, O- and C-rich models with enhanced scattering both find reasonable fits to the STIS, GMOS, and WFC3 data under a variety of conditions.  The {\em Spitzer} data suggest shallower transit depths than the models predict at infrared wavelengths, albeit at low statistical significance.  More precise data are needed over a broad wavelength range to distinguish between the assortment of models presented in this paper.

Future analyses of WASP-12b would benefit from more precise constraints on the absolute pressure level (by measuring the Rayleigh scattering slope in the blue optical) and from non-overlapping spectroscopic observations of WASP-12's stellar companion to obtain a high-precision correction.  The latter is possible with new {\em HST/WFC3} spectroscopic observations at a more optimal role angle.  In a future paper, we will present a reanalysis of previously published dayside emission observations with corrected eclipse depths to complement this work.  To this end, a comprehensive analysis considering both the transmission and emission spectra will provide the best constraints on WASP-12b's composition, thermal profile, and C/O.  As we have shown here, there is a lot to gain from combining ground- and space-based transit observations with broad spectral coverage, but we need additional high-precision data to continue investigating the nature of exoplanetary atmospheres.

\acknowledgments

We thank Carlo Graziani for providing comments related to the {\tt Divide-White} technique and also Jonathan Tennyson and Christian Hill for helpful discussions.  We thank contributors to SciPy, Matplotlib, and the Python Programming Language, the free and open-source community, the NASA Astrophysics Data System, and the JPL Solar System Dynamics group for software and services.  This research made use of Tiny Tim/Spitzer, developed by John Krist for the Spitzer Science Center. The Center is managed by the California Institute of Technology under a contract with NASA.

Funding for this work has been provided by NASA grant NNX13AJ16G. J.L.B. acknowledges support from the Alfred P.~Sloan Foundation. J.-M.D. acknowledges funding from NASA through the Sagan Exoplanet Fellowship program administered by the NASA Exoplanet Science Institute (NExScI).  N.M. acknowledges support from the Yale Center for Astronomy and Astrophysics (YCAA) through the YCAA postdoctoral prize fellowship.  L.K. acknowledges support from the National Science Foundation Graduate Research Fellowship Program.  D.H. acknowledges support from the European Research Council under the European Community's Seventh Framework Programme (FP7/2007-2013 Grant Agreement no. 247060).

This work is based on observations obtained at the Gemini Observatory, which is operated by the 
Association of Universities for Research in Astronomy, Inc., under a cooperative agreement 
with the NSF on behalf of the Gemini partnership: the National Science Foundation 
(United States), the National Research Council (Canada), CONICYT (Chile), the Australian 
Research Council (Australia), Minist\'{e}rio da Ci\^{e}ncia, Tecnologia e Inova\c{c}\~{a}o 
(Brazil) and Ministerio de Ciencia, Tecnolog\'{i}a e Innovaci\'{o}n Productiva (Argentina).
\\

\bibliography{ms}

\end{document}